\documentclass[sigconf]{acmart}

\usepackage{xcolor}
\definecolor{flatgreen}{HTML}{b7f4d8}
\definecolor{flatred}{HTML}{ffcd02}
\definecolor{lightgray}{gray}{0.9}
\definecolor{nolan_green}{HTML}{119F57}
\definecolor{nolan_yellow}{HTML}{F7AB00}
\definecolor{nolan_red}{HTML}{E74C3C}

\newcommand{\edit}[1]{\textcolor{black}{#1}}

\usepackage{xspace}
\usepackage{comment}
\usepackage{listings}
\usepackage{colortbl}
\usepackage{soul}
\usepackage{ifthen}
\usepackage{fontawesome}
\usepackage{array,multirow,graphicx}
\usepackage{float}
\usepackage{pifont}
\usepackage{balance}
\usepackage[most]{tcolorbox}
\usepackage{amsmath}
\usepackage{framed}
\usepackage{siunitx}
\usepackage{bm}
\usepackage{booktabs}
\usepackage{enumitem}
\usepackage{tipa} %

\usepackage{tikz}
\usepackage[normalem]{ulem}
\usepackage{marginnote}
\usepackage[colorinlistoftodos]{todonotes}

\usepackage[utf8]{inputenc}

\definecolor{headergray}{RGB}{240, 242, 245} %
\definecolor{sectgray}{RGB}{225, 228, 232}   %
\definecolor{nolan_green}{RGB}{39, 174, 96}  %
\definecolor{nolan_yellow}{RGB}{243, 156, 18} %

\newcommand{\BadgeC}{\tikz[baseline=(char.base)]{\node[fill=blue!10, text=blue!60!black, inner sep=2pt, rounded corners=2pt, font=\sffamily\bfseries\small] (char) {C};}}
\newcommand{\BadgeE}{\tikz[baseline=(char.base)]{\node[fill=red!10, text=red!60!black, inner sep=2pt, rounded corners=2pt, font=\sffamily\bfseries\small] (char) {E};}}
\newcommand{\BadgeD}{\tikz[baseline=(char.base)]{\node[fill=orange!10, text=orange!60!black, inner sep=2pt, rounded corners=2pt, font=\sffamily\bfseries\small] (char) {D};}}
\newcommand{\BadgeS}{\tikz[baseline=(char.base)]{\node[fill=purple!10, text=purple!60!black, inner sep=2pt, rounded corners=2pt, font=\sffamily\bfseries\small] (char) {S};}}
\newcommand{\BadgeR}{\tikz[baseline=(char.base)]{\node[fill=teal!10, text=teal!60!black, inner sep=2pt, rounded corners=2pt, font=\sffamily\bfseries\small] (char) {R};}}
\newcommand{\BadgeX}{\tikz[baseline=(char.base)]{\node[fill=cyan!10, text=cyan!60!black, inner sep=2pt, rounded corners=2pt, font=\sffamily\bfseries\small] (char) {X};}}
\newcommand{\BadgeF}{\tikz[baseline=(char.base)]{\node[fill=magenta!10, text=magenta!60!black, inner sep=2pt, rounded corners=2pt, font=\sffamily\bfseries\small] (char) {F};}}
\newcommand{\BadgeP}{\tikz[baseline=(char.base)]{\node[fill=brown!10, text=brown!60!black, inner sep=2pt, rounded corners=2pt, font=\sffamily\bfseries\small] (char) {P};}}
\newcommand{\BadgeM}{\tikz[baseline=(char.base)]{\node[fill=gray!10, text=gray!60!black, inner sep=2pt, rounded corners=2pt, font=\sffamily\bfseries\small] (char) {M};}}

\newcommand{\qarating}[2]{P#1 (\textbf{\texttt{#2}})}
\usepackage{xparse}
\usepackage{pgffor}
\usepackage[htt]{hyphenat}
\presetkeys%
    {todonotes}%
    {size=\tiny}
    {}

\usepackage{microtype}
\usepackage{tabularx}
\usepackage{ragged2e}

\newcolumntype{Y}{>{\RaggedRight\arraybackslash}X}
\newcolumntype{Z}{>{\RaggedRight\arraybackslash}p{3cm}}

\definecolor{mGreen}{rgb}{0,0.6,0}
\definecolor{mGray}{rgb}{0.5,0.5,0.5}
\definecolor{mPurple}{rgb}{0.58,0,0.82}
\definecolor{backgroundColour}{rgb}{0.95,0.95,0.92}

\lstdefinestyle{CStyle}{
    backgroundcolor=\color{backgroundColour},
    commentstyle=\color{mGreen},
    keywordstyle=\color{magenta},
    numberstyle=\tiny\color{mGray},
    stringstyle=\color{mPurple},
    basicstyle=\footnotesize,
    breakatwhitespace=false,
    breaklines=true,
    captionpos=b,
    keepspaces=true,
    numbers=left,
    numbersep=5pt,
    showspaces=false,
    showstringspaces=false,
    showtabs=false,
    tabsize=2,
    language=C
}
\usepackage{afterpage}
\usepackage[ruled,linesnumbered]{algorithm2e}
\usepackage{url}
\usepackage{subfigure}
\SetAlFnt{\small}
\SetAlCapFnt{\small}
\SetAlCapNameFnt{\small}
\usepackage{algorithmic}
\algsetup{linenosize=\small}
\SetKwInput{KwInput}{Input}                %
\SetKwInput{KwOutput}{Output}              %

\newboolean{showcomments}
\setboolean{showcomments}{true}
\ifthenelse{\boolean{showcomments}}
{ }

\ifthenelse{\boolean{showcomments}}
{ }

\ifthenelse{\boolean{showcomments}}
{ }

\newtcolorbox{findingbox}{
    boxrule=0pt,
    frame hidden,
    sharp corners,
    enhanced,
    borderline north={1pt}{0pt}{black},
    borderline south={1pt}{0pt}{black},
    boxsep=2pt,
    left=2pt,
    right=2pt,
    top=2.5pt,
    bottom=2pt,
    before skip=3pt plus 1pt minus 1pt,
    after skip=3pt plus 1pt minus 1pt
}

\tcbset {
  base/.style={
    arc=0mm, 
    bottomtitle=0.5mm,
    boxrule=0mm,
    colbacktitle=black!10!white, 
    coltitle=black, 
    fonttitle=\bfseries, 
    left=2.5mm,
    leftrule=1mm,
    right=3.5mm,
    title={#1},
    toptitle=0.75mm, 
  }
}

\definecolor{brandblue}{rgb}{0.34, 0.7, 1}

\newtcbox{\mainbadge}{on line,
  tcbox raise base,
  height=2.3ex,             %
  valign=center,            %
  colback=brandblue!85!black,
  colframe=brandblue!85!black,
  coltext=white,
  fontupper=\sffamily\mdseries,
  boxrule=.5pt,
  arc=2pt,
  left=0pt,
  right=0pt,
  baseline=0pt,
  nobeforeafter}

\newtcbox{\sidebadge}{on line,
  tcbox raise base,
  height=2.3ex,             %
  valign=center,
  colback=white,
  colframe=black!35,
  coltext=black,
  fontupper=\sffamily\mdseries,
  boxrule=.5pt,
  arc=2pt,
  left=0pt,
  right=0pt,
  baseline=0pt,
  nobeforeafter}

\newcommand{\showbadges}[2]{%
  \if\relax\detokenize{#1}\relax\else
    \foreach \x in {#1}{\mainbadge{\x}\hspace{1pt}}%
  \fi
  \if\relax\detokenize{#2}\relax\else
    \foreach \x in {#2}{\sidebadge{\x}\hspace{1pt}}%
  \fi
}

\NewDocumentEnvironment{challengebox}{ m O{} O{} }{%
  \begin{tcolorbox}[
    enhanced,
    colframe=brandblue,
    base={#1},
    boxsep=1pt,
    top=0pt,
    bottom=0pt,
    before skip=0pt,
    after skip=0pt,
    enlarge top by=3pt,
    enlarge bottom by=3pt,
    before upper={\vspace{-2pt}\noindent\showbadges{#2}{#3}\enspace\ignorespaces}
  ]%
}{%
  \end{tcolorbox}%
}

\newtcolorbox{subbox}[1]{
  colframe=black!30!white,
  base={#1}
}

\definecolor{RoyalBlue}{RGB}{65,105,225}

\AtBeginDocument{%
  \providecommand\BibTeX{{%
    \normalfont B\kern-0.5em{\scshape i\kern-0.25em b}\kern-0.8em\TeX}}}

\acmConference[EASE 2026]{The 30th International Conference on Evaluation and Assessment in Software Engineering}{9--12 June, 2026}{Glasgow, Scotland, United Kingdom}

\begin{document}
\title{Industry Practitioners' Perspectives on AI Model Quality: Perceptions, Challenges, and Solutions}

\author{Chenyu Wang}
\email{chenyuwang@smu.edu.sg}
\affiliation{%
  \institution{Singapore Management University}
  \city{Singapore}
  \country{Singapore}
}

\author{Zhou Yang}
\email{zhou.yang@ualberta.ca}
\affiliation{%
  \institution{University of Alberta}
  \city{Edmonton}
  \state{Alberta}
  \country{Canada}
}

\author{Yunbo Lyu}
\authornote{Corresponding author}
\email{yunbolyu@smu.edu.sg}
\affiliation{%
  \institution{Singapore Management University}
  \city{Singapore}
  \country{Singapore}
}

\author{Ze Shi Li}
\email{zeshili@ou.edu}
\affiliation{%
  \institution{University of Oklahoma}
  \city{Norman}
  \state{Oklahoma}
  \country{United States}
}

\author{Daniela Damian}
\email{danielad@uvic.ca}
\affiliation{%
  \institution{University of Victoria}
  \city{Victoria}
  \state{British Columbia}
  \country{Canada}
}

\author{David Lo}
\email{davidlo@smu.edu.sg}
\affiliation{%
  \institution{Singapore Management University}
  \city{Singapore}
  \country{Singapore}
}

\renewcommand{\shortauthors}{Wang et al.}

\begin{abstract}
  Artificial Intelligence (AI) is ubiquitous, powering applications across nearly every industry.
  With this broad adoption, assuring AI model quality is essential for building reliable and trustworthy systems.
  Historically, correctness has been the primary focus, yet industry AI models require many other critical quality attributes.
  To understand industry perceptions, challenges, and solutions regarding these attributes, we identify nine key quality attributes and interview 15 AI practitioners from diverse backgrounds.
  Our interviews reveal that practitioners' perceptions vary; for example, \textit{efficiency} is prioritized over \textit{correctness} for real-time applications, and \textit{scalability} and \textit{deployability} are no longer primary concerns.
  Data imbalance stands out as a major obstacle to maintaining model \textit{correctness} and \textit{robustness}; accordingly, practitioners employ mitigation strategies like active learning.
  We validate our key findings via a survey of 50 practitioners, showing most findings are well-acknowledged.
  These insights can guide researchers to prioritize attributes valued by practitioners, avoiding approaches that improve one attribute at the expense of others deemed more critical.
\end{abstract}



\maketitle

\section{Introduction}
\label{sec:introduction}
In recent decades, Artificial Intelligence (AI) technologies have rapidly advanced across critical domains, such as finance~\cite{bao2022artificial} and image processing~\cite{goodfellow2014generative}.
However, like traditional software, AI models are prone to bugs, failures, and quality issues that can lead to severe consequences.
As reported in 2022, nearly 400 car crashes in 11 months involved AI-based automated technologies~\cite{Press_2022}.
Wong et al.~\cite{wong2021external} find that a widely-used sepsis identification AI misses most cases and raises frequent false alarms, and Chen et al.~\cite{chen2025secureagentbench} show that LLM code agents may produce insecure code.
Ensuring AI model quality is more urgent and important than ever before.

Unlike traditional software that operates through explicit, deterministic logic with predictable control flow, most AI models are inherently data-driven systems that learn patterns through training processes and often function as black boxes.
Consequently, the numerous well-established principles and practices for ensuring quality in traditional software development may not directly apply to AI model development, whose bugs often stem from data rather than code---rendering techniques like tracing bug-inducing commits~\cite{lyu2024evaluating} ill-suited.
Therefore, the field of AI model quality requires explicit and dedicated study.

Previous academic research has explored various aspects of AI quality, such as correctness~\cite{yang2022revisiting} and robustness (preventing mistakes on unexpected inputs)~\cite{acsac2022gong}.
However, practices and methods in academia-based research papers may not be easily adopted by industry practitioners.
AI models in the industry are usually larger and more complex than the ones evaluated in research papers~\cite{song2022exploring}.
Furthermore, standards and regulations that industry AI models need to meet are usually many-sided and domain-specific.
Such differences call for a direct analysis from industry practitioners' perspectives.
Although prior studies have examined challenges and practices in AI development~\cite{10.1145/3533378, 10.1145/3382494.3410681} under an industry context, they focus primarily on process-level activities (e.g., peer review, team collaboration, etc.) and leave unanswered how actions on these activities translate into concrete quality attributes, such as correctness and fairness that correlate with business success.
Understanding which quality attributes practitioners prioritize and why helps clarify real-world needs and guides focus on what matters most in different AI contexts.
Identifying challenges and solutions provides practical guidance and reveals directions for future research.

Aiming to fill this gap, we conduct a series of interviews with fifteen AI experts working in the industry.
We identify nine key AI quality attributes, i.e., \textit{correctness}, \textit{robustness}, \textit{efficiency}, \textit{fairness}, \textit{explainability}, \textit{privacy}, \textit{deployability}, \textit{scalability} and \textit{maintainability}.
We interview practitioners to understand their perceptions of the importance of each quality attribute (RQ1) and to identify the specific challenges they face and solutions they employ for each quality attribute (RQ2).
Four perceptions and four key challenges have been identified, along with corresponding solutions.
We then validate our findings through a survey of fifty industry practitioners (RQ3), seven out of eight findings are well acknowledged, and the remaining one is marginally agreeable.

Our findings provide insights that are overlooked by prior surveys about perceptions, challenges, and solutions of AI development.
For example, regarding practitioners’ perceptions, we observe a shift in how \textit{deployability} and \textit{scalability} are prioritized.
While earlier work frames these as key quality attributes that AI engineers should emphasize~\cite{10.1145/3533378, nahar2023meta}, practitioners in our study report that such concerns are increasingly delegated to platform infrastructure.
Regarding practitioners’ challenges and solutions, we notice that to address data imbalance, active learning is effective for collecting new representative data, and for the need of data synthesis, practitioners still stick to rule-based traditional augmentation and resampling methods.

\section{Background}
\label{sec:background}

\subsection{AI Models}
\label{sec:ml_models}
Artificial Intelligence (AI) \textit{`seeks to make computers do the sorts of things that minds can do'}~\cite{boden2016ai}.
AI models employ diverse techniques across domains, like reinforcement learning for recommendations~\cite{xiangyu2017deep} and natural language processing for summarization~\cite{widyassari2022review}.
Unlike traditional explicit software, AI models are data-driven, learning from training data rather than predefined rules, which introduces unique perceptions and challenges.
This study covers a broad range of AI models, encompassing traditional machine learning, deep neural networks, and large language models.

\subsection{Quality Attributes of AI Models}
\label{sec:quality_attributes_of_ml_systems}
Quality attributes are characteristics that define how well a system performs its functions and meets stakeholder expectations~\cite{iso25010-2011}, they provide a valuable analytical framework that helps identify perceptions and challenges beyond accuracy while enabling systematic comparison across projects and domains.
As a shared vocabulary, they also facilitate discussions and comparisons between stakeholders and domains.
For example, both recommender and video streaming systems require high \textit{efficiency} for real-time response.

While AI models share the same quality attributes as traditional software, they fundamentally alter the mechanisms that influence them.
A prime example is \textit{fairness}: in traditional software, bias usually stems from explicit logic or assumptions made during design.
In contrast, AI models introduce a new vector for bias through the data they ingest.
Since training data often encodes historical or societal prejudices, AI systems can unintentionally reproduce or amplify unfair outcomes, even without biased code.

\section{Research Design}
\label{sec:methodology}

We adopt a mixed-methods design: semi-structured interviews with 15 practitioners elicit in-depth accounts addressing RQ1 and RQ2, and a follow-up survey with 50 practitioners validates these findings at scale (RQ3).

\noindent\textbf{RQ1: How do AI practitioners perceive quality attributes?}
Industry resources are finite, practitioners should focus on attributes that best serve the domain and business objectives.
Clarifying how they prioritize these attributes and under which conditions each one becomes critical will both sharpen industrial practices and steer academic research toward the qualities that matter most in real-world settings.

\vspace{0.1cm}
\noindent\textbf{RQ2: What are the challenges and solutions in ensuring different quality attributes?}
Identifying attribute-specific challenges unveils industry pain points and highlights actionable research directions.
Uncovering real-world solutions practitioners adopted to address these challenges can offer valuable insights for others facing similar issues.

\vspace{0.1cm}
\noindent
\textbf{RQ3: To what extent are our findings recognized in practice?}
This research question examines whether and to what degree the perceptions, challenges, and solutions identified in this study align with the experiences and judgments of industry practitioners.

\begin{table}[t]
  \footnotesize
  \centering
  \caption{Practitioners' index, job title, task of expertise, years of AI experience, and company size (number of employees).}
  \vspace{-1em}
  \label{tab:participants}
  \renewcommand{\arraystretch}{1.15}
  \begin{tabularx}{\linewidth}{l l X c l}
    \toprule
    \textbf{Idx} & \textbf{Job Title} & \textbf{Task of Expertise} & \textbf{Exp} & \textbf{\#Emp.} \\
    \midrule
    P1  & ML Engineer              & Fake Document Detection                        & 2 & 10K+         \\
    P2  & ML Research Engineer     & Vulnerability Detection                   & 4 & 10K+         \\
    P3  & ML Algorithm Engineer    & Financial Fraud Detection                      & 6 & 100--1K      \\
    P4  & GPU Software Engineer                  & Algorithm Validation                           & 7 & 10K+         \\
    P5  & Senior Data Engineer                   & Data Integration                 & 3 & 100--1K      \\
    P6  & AI Algorithm Engineer                  & AI for Financial Fraud                   & 5 & 100--1K      \\
    P7  & Data Scientist                         & Video Streaming                                & 2 & 100--1K      \\
    P8  & Senior Data Scientist                  & Recommendation System                          & 3 & 10K+         \\
    P9  & ML Engineer              & General AI Algorithm                           & 4 & 10K+         \\
    P10 & AI Engineer                            & BI Provider                 & 5 & 100--1K      \\
    P11 & AI Applied Scientist                   & Ads, e-Commerce                                & 6 & 10K+         \\
    P12 & AI Engineer                            & Financial Risk Control                         & 5 & 1K--10K      \\
    P13 & CV Researcher             & Synthetic Media Detection                      & 2 & 1K--10K      \\
    P14 & NLP Engineer   & Text Summarization                             & 2 & 10--100      \\
    P15 & NLP Researcher & Code Clone, Bug Fixing                         & 3 & 1K--10K      \\
    \bottomrule
  \end{tabularx}
\end{table}

\subsection{Participant Selection}
\label{sec:participant_selection}
We recruit AI professionals with at least 2 years of industry experience, covering various roles.
The initial pool of participants is derived from our personal contacts within various companies.
To augment this group, we extend our search to referrals from our initial interviewees using snowball sampling~\cite{goodman1961snowball}.
This approach yields a diverse cohort of 15 interviewees, each from a distinct company, with demographics as detailed in Table~\ref{tab:participants}.
Their experience ranges from 2 to 7 years (median = 4), spanning five countries (Singapore, China, Canada, the United States, and Australia), companies from startups to multinational enterprises, and diverse AI-related roles (e.g., data engineers, algorithm engineers, industry-based researchers) across domains such as finance, information technology, and e-commerce.

\subsection{Attribute Selection}
\label{subsec:attributes_selection}

To identify the quality attributes that industry practitioners care about, we follow a three-step process.
First, we review prior literature to compile an initial attribute set.
Second, we validate and refine this set through interviews with practitioners.
Third, we further assess and refine its relevance using a questionnaire survey administered to a broader group of industry practitioners.

We begin by reviewing existing quality models and taxonomies in the literature.
Comprehensive frameworks, such as ISO/IEC 25059~\cite{iso25059-2023} and non-functional requirement taxonomies for ML systems~\cite{habibullah2022non}, provide fine-grained lists comprising over 30 quality attributes.
However, their level of detail and conceptual overlap (e.g., between explainability and interpretability) makes them impractical for time-constrained interviews.
To balance focus and coverage, we adopt Zhang et al.'s taxonomy~\cite{zhang2019machine} as our baseline: unlike the broader frameworks, it is derived specifically from ML testing literature and consolidates quality concerns into seven properties at a granularity tractable for time-boxed interviews.
Two authors jointly map attributes from ISO/IEC 25059~\cite{iso25059-2023} and Habibullah et al.~\cite{habibullah2022non} onto these properties and iteratively resolve disagreements.
This mapping indicates that most fine-grained attributes can be subsumed under the seven properties without loss of meaning.
Based on this process, we refine the taxonomy by (1) excluding \textit{model relevance}, which primarily reflects overfitting and is therefore subsumed under \textit{robustness}, and (2) renaming \textit{interpretability} to \textit{explainability} to align with industrial terminology.

We further validate and refine the attribute set through interviews (Section~\ref{subsec:interview_scheme}) with experienced industry practitioners.
During interviews, we consider whether to separate \textit{robustness} and \textit{security} but ultimately retain a combined definition, as practitioners view them as jointly capturing a model's resilience under adversarial or anomalous conditions.
Based on practitioner feedback, we additionally include two attributes that are particularly salient in industrial settings: \textit{scalability}~\cite{Barmer2021} and \textit{deployability}~\cite{10.1145/3533378}.
To assess the broader relevance of the refined attribute set, we validate it through a survey (Section~\ref{subsec:survey}).
Based on the survey results, the majority of respondents (43 out of 50) agree that the attribute set comprehensively captures their key quality concerns without redundancy.
Following suggestions from seven respondents, we further review our interview content and find that maintainability is indeed a concern; accordingly, we incorporate \textit{maintainability}, scoped specifically to operational concerns such as model upgrading and monitoring.
Implementation-level aspects (e.g., code quality) are excluded, as they pertain to the software codebase rather than model-level behavioral quality.

This process results in nine quality attributes for our study: \textit{correctness}, \textit{robustness}, \textit{efficiency}, \textit{fairness}, \textit{explainability}, \textit{privacy}, \textit{scalability}, \textit{deployability}, and \textit{maintainability}.
Definitions of these attributes are provided in Section~\ref{sec:rq1}.
During the main interviews, all 15 participants confirm that this set adequately covers their key quality concerns.

\subsection{Interview}
\label{subsec:interview_scheme}

\edit{
We conduct 45--70 minute semi-structured interviews to explore practitioners' perceptions, challenges, and solutions regarding the nine quality attributes identified in Section~\ref{subsec:attributes_selection}. 
During the interviews, we use slides to clarify key terms, such as the definitions of quality attributes, to ensure a shared understanding among participants.
We formulate our research questions based on existing literature about quality attributes of AI models (Section~\ref{sec:related_work}). 
After conducting two pilot interviews and making revisions, we finalize an interview protocol consisting of 40 questions, which are detailed in our replication package. 
The pilot study further confirms that the selected nine quality attributes are reasonable and relevant in industry contexts.
}
During the interviews, we first ask participants to confirm or refine whether the nine attributes adequately represent the quality concerns in their application domains.
If confirmed, we then ask participants to rate the importance of each attribute within their domain and to justify their ratings. Importance is assessed using a 5-point scale: \texttt{5}—critically important, \texttt{4}—highly important, \texttt{3}—moderately important, \texttt{2}—minimally important, and \texttt{1}—not important.
These ratings and justifications allow us to examine which quality attributes practitioners prioritize across different domains and contexts, thereby addressing RQ1.

\edit{
We further ask participants to describe the specific challenges they encounter when ensuring each quality attribute, as well as the strategies they adopt to mitigate these challenges, to address RQ2.
In addition to attribute-centric questions, we also prompt participants to discuss issues arising at different stages of the AI development lifecycle. We subsequently map these stage-specific issues to quality attributes during analysis.
This design helps reduce recall bias and captures challenges that practitioners may not explicitly associate with particular quality attributes.
}

\subsection{Analysis of Interview}
\label{subsec:analysis_of_interview}

\edit{
For each interview, we record the audio and generate transcripts using speech-to-text services (e.g., Zoom).
Two authors then proofread and correct all transcripts against the recordings together to improve transcription accuracy.
After transcribing the interviews, we analyze the transcripts following thematic synthesis~\cite{10.1109/ESEM.2011.36}.
As discussed in Section~\ref{subsec:attributes_selection}, participants generally confirm that the nine attributes adequately cover their major quality concerns. 
We therefore adopt a hybrid coding approach, combining deductive coding with inductive thematic analysis.
}

For deductive coding, we use two predefined code sets: (1) the nine quality attributes, and (2) three aspect codes (\textit{perception}, \textit{challenge}, and \textit{solution}).
While participants generally agree that the nine attributes capture their main quality concerns, we allow out-of-framework coding for segments that do not fit.
In our data, Other Attributes segments are rare and heterogeneous and do not form a coherent pattern, so we focus our analysis on the nine attributes.
We segment the transcripts into minimal meaning units, splitting a segment whenever the speaker shifts topic or discourse function (e.g., from perception to challenge/solution), and conduct coding at the segment level.
Each segment is assigned codes from both dimensions, forming an attribute-aspect pair.
For instance, when a participant describes a perception related to scalability, we code the segment as \textit{scalability}~$\times$~\textit{perception}.
We also allow cross-attribute coding when the same segment relates multiple attributes.
For example, if a participant discusses how imbalanced training data affects both model accuracy and robustness under distribution shifts, we assign both \textit{correctness} and \textit{robustness} codes to that segment.
Two authors independently code all transcripts.
After completing the initial coding, they meet to compare results and resolve disagreements through discussion, achieving a Cohen's kappa of 0.76, indicating substantial inter-rater agreement~\cite{landis1977measurement}.

Following the deductive coding, we conduct inductive thematic analysis within each attribute-aspect group.
Two authors independently review all segments within each group and propose emergent themes.
For example, within the \textit{scalability}~$\times$~\textit{perception} group, we identify themes such as \textit{Scaling overhead is mostly handled by platform infrastructure}.
The two authors then meet to consolidate themes, merging overlapping ones and refining theme definitions until reaching consensus.
Consequently, we identify 4 perception themes (Section~\ref{sec:rq1}), 4 challenge themes, and 6 corresponding solution themes (Section~\ref{sec:rq2}).
After drafting the manuscript, we conduct member checking by inviting all participants to review our findings.
Five participants respond and confirm that our interpretations accurately represent their experiences.

\vspace{-0.3em}
\subsection{Survey}
\label{subsec:survey}

To validate the perceptions, challenges, and solutions derived from our interviews, we conduct a follow-up survey targeting a broader population of AI practitioners.
The survey consists of two parts.
The first part collects demographic information, including years of AI experience in industry, current role, and application domain.
The second part presents statements distilled from our interview findings, covering perceptions (Section~\ref{sec:rq1}), challenges, and solutions (Section~\ref{sec:rq2}).
For each statement, respondents indicate their level of agreement on a 5-point Likert scale (1 = strongly disagree, 5 = strongly agree), with an additional ``I don't know'' option for cases where the statement does not apply to their context.
Each item also includes an optional free-text field for elaboration.
The full questionnaire is available in our replication package.

We recruit 61 practitioners via Prolific, consistent with prior empirical studies that utilize crowdsourcing platforms to engage AI professionals~\cite{baldwin2025ethicspracticesaidevelopment}.
To ensure our sample reflects industrial AI development experience, we exclude 11 respondents identified as general software engineers or IT professionals, resulting in 50 valid responses.
Our participants originate from North America, Asia, and Europe, spanning diverse roles such as ML/AI engineers, data scientists, and applied researchers.
The cohort possesses a median of 3 years of experience (range: 1--8 years) across varied application domains, such as finance, computer vision, LLM-based chatbots, and recommendation systems.

\vspace{-0.6em}

\section{Results}
\label{sec:results}
In this section, we present the perceptions (RQ1), challenges and solutions (RQ2) extracted from the interviews and compare them with the related literature.
We further validate these insights through a follow-up survey (RQ3).

\subsection{RQ1: How do AI Practitioners Perceive Quality Attributes?}
\label{sec:rq1}

\edit{
We first present the nine quality attributes, along with the definitions synthesized from the literature and the pilot study described in Section~\ref{subsec:attributes_selection}.}

\noindent
\edit{\textbf{\textit{Correctness}} evaluates how reliably the model performs its intended function and produces the expected outputs.}

\noindent
\edit{\textbf{\textit{Robustness}} refers to the model's ability to produce reliable and benign results when exposed to new, unseen, or noisy data and resist adversarial input perturbations. 
Following Zhang et al.~\cite{zhang2019machine}, we recognize that robustness failures are often the root cause of security vulnerabilities (e.g., adversarial attacks) and safety risks.
Our interviews did not reveal security or safety issues that were both absent from prior surveys and incapable of being subsumed under robustness.
Consequently, rather than establishing separate attributes, we subsume these concerns under the umbrella of \textit{Robustness} for conciseness of presentation.
}

\noindent
\edit{\textbf{\textit{Efficiency}} denotes a trained AI model's ability to produce outputs with minimal time, computational power, and energy during inference.
While training efficiency is also important, our study focuses on the quality of the trained AI model itself rather than the training process, so we do not consider training efficiency.}

\noindent
\edit{\textbf{\textit{Fairness}} refers to the extent to which a model avoids producing unjustified disparities or discriminatory outcomes across different demographic groups or protected attributes.}

\noindent
\edit{\textbf{\textit{Explainability}} (often closely related to \textit{interpretability}) refers to the ability to explain a model's decision-making processes.
It enables humans to understand the factors that contribute to the model's decisions.}

\noindent
\edit{\textbf{\textit{Privacy}} concerns the extent to which personal and sensitive data related to the AI model (e.g., training data) is protected from unauthorized access.} 

\noindent
\edit{\textbf{\textit{Scalability}} measures the model's ability to maintain acceptable performance when its workload or data volume increases.}

\noindent
\edit{\textbf{\textit{Deployability}} measures the ease with which an AI model can be integrated into existing environments, workflows, or systems while maintaining its performance and functionality.}

\noindent
\edit{\textbf{\textit{Maintainability}} refers to the ease with which an AI model can be updated, debugged, and improved over time. We focus on operational concerns like model upgrading and monitoring in this study. Implementation-level issues (e.g., code quality) are not discussed.}

During the interviews, participants rated the importance of each quality attribute on a five-point Likert scale based on their professional experience.
\textit{Correctness} emerges as the highest-priority attribute, with 11 out of 15 participants rating it as critically important (5) and the remaining 4 as very important (4).
\textit{Efficiency} and \textit{robustness} are similarly emphasized: 11 and 12 participants, respectively, rate these attributes as either critically important (5) or very important (4).
\textit{Maintainability} is also considered important, with 9 participants assigning it a rating of 4 or 5.
By contrast, \textit{fairness} receives comparatively less attention in industrial practice, as 10 out of 15 participants rate it as not important (1) or only slightly important (2).
The remaining attributes—\textit{explainability}, \textit{privacy}, \textit{deployability}, and \textit{scalability}—exhibit more heterogeneous assessments, with ratings spanning the full range of the scale.

\edit{  
Following are four key perceptions regarding quality attributes that emerged from our interviews.
To ensure brevity, practitioners' opinions are quoted with their ratings, e.g., \qarating{3}{5} indicates that respondent P3 assigned a rating of \textbf{\texttt{5}} to the attribute being discussed at the time.
The summary of findings is presented in Table~\ref{tab:takeaways}.
}

\begin{table*}[htbp]
    \small
    \centering
    \caption{Summary of perceptions, challenges, and solutions derived from the interviews, along with the scores they received from the validation survey ($\geq$4: \textcolor{nolan_green}{well-acknowledged}; 3.5--4: \textcolor{nolan_yellow}{marginally agreeable}).
    \textbf{Attributes Key:} 
    \protect\BadgeC~Correctness \enspace 
    \protect\BadgeD~Deployability \enspace 
    \protect\BadgeE~Efficiency \enspace 
    \protect\BadgeF~Fairness \enspace 
    \protect\BadgeM~Maintainability \enspace 
    \protect\BadgeP~Privacy \enspace 
    \protect\BadgeR~Robustness \enspace 
    \protect\BadgeS~Scalability \enspace 
    \protect\BadgeX~Explainability}
    \label{tab:takeaways}
    
    \renewcommand{\arraystretch}{1.2} 
    
    \begin{tabularx}{\textwidth}{@{} c >{\RaggedRight\arraybackslash}X l c @{}}
        \toprule
        \rowcolor{headergray}
        \textbf{\sffamily \#} & \textbf{\sffamily Finding Content} & \textbf{\sffamily Attribute(s)} & \textbf{\sffamily Score} \\
        \midrule
        
        \rowcolor{sectgray}
        \multicolumn{4}{@{} p{\textwidth} @{}}{\hspace{3pt}\textbf{\sffamily \textit{Perceptions}}} \\
        \addlinespace[3pt] 
        
        1 & In real-time AI, operational cost, response latency, and model capability form a \textbf{three-way tradeoff}. Practitioners prioritize \textit{efficiency} by optimizing latency and cost, while ensuring \textit{correctness} remains within acceptable limits. 
          & \BadgeC~\BadgeE 
          & \textcolor{nolan_green}{\textbf{4.16}} \\
        \addlinespace 
        
        2 & While previously considered challenging, \textit{deployability} and \textit{scalability} are now handled by \textbf{microservices}. AI teams containerize models and expose APIs for seamless integration, and utilize platform infrastructure for dynamic scaling. 
          & \BadgeD~\BadgeS 
          & \textcolor{nolan_green}{\textbf{4.02}} \\
        \addlinespace
        
        3 & \textbf{Compliance} translates legal mandates into quality attributes: safety necessitates \textit{robustness} in GenAI, auditability requires \textit{explainability} in finance, anti-discrimination dictates \textit{fairness} for demographics, and data protection enforces \textit{privacy}. 
          & \BadgeR~\BadgeX~\BadgeF~\BadgeP 
          & \textcolor{nolan_green}{\textbf{4.24}} \\
        \addlinespace
        
        4 & Beyond compliance, \textit{explainability} aids \textit{maintainability} and fosters user trust. While industry practitioners used to focus primarily on the former, today they place equal importance on the latter. 
          & \BadgeX~\BadgeM
          & \textcolor{nolan_green}{\textbf{4.18}} \\
        \addlinespace
        
        \midrule
        
        \rowcolor{sectgray} 
        \multicolumn{4}{@{} p{\textwidth} @{}}{\hspace{3pt}\textbf{\sffamily \textit{Key Challenges and Solutions}}} \\
        \addlinespace[3pt]
        
        1 & To mitigate data imbalance, \textbf{active learning} proves effective for data acquisition, while traditional data resampling and augmentation methods prevail in data synthesis due to the control they offer. 
          & \BadgeC~\BadgeR~\BadgeF~\BadgeX
          & \textcolor{nolan_green}{\textbf{4.13}} \\
        \addlinespace
        
        2 & While \textbf{data drift} is a critical concern, monitoring complexities often compel the industry to bypass drift detection in favor of pragmatic retraining based on time or newly collected data volume. 
          & \BadgeC~\BadgeR~\BadgeM
          & \textcolor{nolan_green}{\textbf{4.00}} \\
        \addlinespace
        
        3 & To minimize operational costs while maintaining accuracy and latency, practitioners coordinate \textbf{hardware selection} (CPU-based inference) with \textbf{software optimization} (model compression). 
          & \BadgeC~\BadgeE 
          & \textcolor{nolan_green}{\textbf{4.10}} \\
        \addlinespace
        
        4 & \textbf{AI-assisted annotation} reduces labeling costs and might even surpass human-only quality. 
          & \BadgeC~\BadgeE~\BadgeM
          & \textcolor{nolan_yellow}{\textbf{3.98}} \\
        \bottomrule
    \end{tabularx}
\end{table*}

\subsubsection{\textbf{Real-time applications: efficiency as the dominant constraint on correctness}}
While academic research often prioritizes state-of-the-art (SOTA) predictive accuracy, real-time industrial applications operate under a fundamentally different set of constraints.
In real-time scenarios involving models that serve massive concurrent users, such as information retrieval and recommendation, \textit{efficiency} (latency and cost) often constrains \textit{correctness} (\qarating{7}{5}, \qarating{8}{5}, \qarating{11}{5}).
In this context, speed drives user satisfaction, and cost-efficiency dictates revenue.

The consensus is clear: if an AI feature introduces perceptible lag, it becomes counterproductive regardless of its correctness. 
P8 remarks, \textit{``\ldots\ If the system is not fast enough, no matter how good it is, users won't even have the patience to use it\ldots''}.
This is not merely anecdotal; P11 quantifies the risk, noting that a search response exceeding 1 second triggers a user abandonment rate of over 30\%, substantially diminishing the practical value of improved search accuracy.
Beyond latency, the operational cost of serving a large user base creates a hard ceiling on model complexity.
P8 emphasizes the business reality: \textit{``\ldots\ With turning a profit as a key objective, we prioritize efficient models that are small, cost-effective, and capable of achieving acceptable accuracy.''}

While prior studies identify latency, cost, and correctness as distinct challenges~\cite{nahar2023meta,you2025navigating,10.1145/3533378}, our findings characterize them as a tightly coupled~\textit{tri-objective tradeoff}. 
In industry practice, latency and cost often act as non-negotiable hard constraints.
These constraints shift the Pareto frontier, rendering correctness often the most adjustable component in this three-way balancing act.

For offline models, \textit{efficiency} tends to become a secondary concern (\qarating{1}{2}, \qarating{9}{2}, \qarating{10}{1}).
Since the inference process is isolated from the user request loop, these models are not bound by the strict latency budgets that govern real-time interactions.
Furthermore, the workload often consists of static, aggregate tasks rather than unique, personalized queries, allowing for significant resource amortization.
P10 provides a clear example of this \textit{compute-once, serve-many} paradigm: \textit{``Our BERT-based AI models precompute the business-intelligence reports on a fixed schedule\ldots\ Clients subscribed to the same sector receive identical reports.''}
In this context, latency and cost sensitivities are diminished, as resource expenditures are amortized across the user base and processing is scheduled during off-peak windows.

\begin{findingbox}
    \textbf{Perception 1}: \textit{Efficiency} is a dominant constraint for AI models deployed in real-time settings, shaping how \textit{correctness} can be pursued and optimized.
\end{findingbox}

\subsubsection{\textbf{Microservices: shifting scalability and deployability burdens away from AI developers}}
While previous surveys identify \textit{deployability} and \textit{scalability} as persistent bottlenecks in industrial AI (e.g.,~\cite{10.1145/3533378, nahar2023meta}), our study uncovers a striking divergence.
Practitioners in industrial sectors like e-commerce and finance (e.g., P3, P6, P8, P11) report that these challenges have largely shifted away from the AI developer's radar due to the strategic adoption of microservices~\cite{newman2021building} and serverless architectures.

Interestingly, this architectural preference was not originally driven by AI.
In the financial sector, microservices were adopted early to minimize single points of failure and localize operational risk.
As P3 explains, \textit{``In the financial sector, clients want the smallest possible component dedicated to a single function for easier deployment, maintenance, scaling, and debugging. This isolation makes sure that the failure of one service won’t affect other critical operations.''}

AI has emerged as a primary beneficiary of this architectural paradigm following its widespread industrial adoption.
Because AI engineering and traditional software engineering involve distinct workflows and lifecycles, tightly coupling them often leads to friction and high debugging costs during deployment.
Microservices solve this by enforcing isolation.
AI components can be easily containerized and deployed using platforms such as NVIDIA Triton Inference Server~\cite{triton}.
This enables AI engineers to expose models via standardized API contracts, allowing collaborating software engineers to consume intelligence as a service, substantially reducing the need to spend time integrating and deploying models alongside other software components.
\qarating{3}{2}, \qarating{6}{1}, \qarating{8}{1}, and \qarating{11}{2} report experiencing little to no pain in terms of \textit{deployability}.

Furthermore, scaling becomes an infrastructure task rather than a code problem: demand spikes simply trigger the auto-scaling of the model-serving container (e.g., via AWS or Azure as described by P8 and P11), leaving the surrounding system untouched.
Therefore, with respect to \textit{scalability}, respondents corresponding to \qarating{3}{1}, \qarating{6}{1}, \qarating{8}{1}, and \qarating{11}{2} indicate that it no longer needs to be a primary concern.
This finding suggests that in mature organizations, \textit{deployability} and \textit{scalability} have been \emph{infrastructuralized}—absorbed by platform capabilities rather than solved by individual developers.
This shifts the AI developer's focus away from `plumbing' mechanics and back toward model quality and business value.

\begin{findingbox}
    \textbf{Perception 2}: Microservice architectures largely shift \textit{deployability} and \textit{scalability} concerns away from individual AI developers. 
    AI teams primarily focus on containerizing models and exposing standardized APIs, while integration, deployment, and dynamic scaling are handled at the infrastructure level.
\end{findingbox}

\subsubsection{\textbf{Compliance as the legislative specification for quality}}
Compliance acts as more than just a constraint; practitioners perceive it as a rigid specification that dictates architectural priorities.
Crucially, we find that compliance often compels industrial AI models to prioritize specific quality attributes, though the precise selection is contingent on the application domain. 
Besides attributes like \textit{fairness} and \textit{privacy} that are intuitively linked to legal ethics, we find that specific regulations also strictly mandate \textit{robustness} and \textit{explainability} in varying domains.
Unlike prior surveys that view compliance broadly as a challenge~\cite{nahar2023meta,10.1145/3533378}, our findings map specific regulatory ``rules'' directly to different prioritized quality attributes.

\vspace*{0.1cm}
\noindent\textbf{The Rule of Safety $\rightarrow$ \textit{Robustness}.}
Regulations like the EU AI Act~\cite{euaiact2024} explicitly forbid AI from generating illegal or harmful content (e.g., bomb-making instructions).
This legal ``red line'' forces generative models to prioritize \textit{robustness}.
As \qarating{3}{5} warns, compliance is not just about typical errors; the model must be robust enough to withstand ``jailbreaking'' attempts and malicious prompts without slipping into non-compliant behavior.

\vspace*{0.1cm}
\noindent\textbf{The Rule of Auditability $\rightarrow$ \textit{Explainability}.}
In highly regulated sectors like finance, ``black box'' decisions are legally unacceptable. 
Regulations (e.g.,~\cite{occ_sar,eu_aml}) mandate that adverse actions—such as freezing suspicious transactions—must be traceable and justifiable to regulators.
Consequently, \textit{explainability} becomes a compliance necessity.
Practitioners (\qarating{3}{5}, \qarating{6}{5}) emphasize that without the ability to explain \textit{why} a transaction was flagged, the system fails to meet the transparency standards required for auditing~\cite{LIN2022118354}.

\vspace*{0.1cm}
\noindent\textbf{The Rule of Non-Discrimination $\rightarrow$ \textit{Fairness}.}
Anti-discrimination laws are the direct drivers for \textit{fairness}.
For high-risk applications like credit scoring, \qarating{6}{5} notes the direct legal threat: \textit{``We face legal action if women receive lower credit scores than men with similar financial histories.''}
Even in lower-risk areas like recommendation systems, strictly ``neutral'' algorithms can still violate laws if they produce disparate impacts based on race or skin color, as feared by \qarating{11}{5}.
Similar concerns extend to generative models~\cite{lyu2025existing}.
Here, compliance transforms \textit{fairness} from a social good into a quantitative legal requirement when demographic information is incorporated into the model's predictions.

\vspace*{0.1cm}
\noindent\textbf{The Rule of Data Sovereignty $\rightarrow$ \textit{Privacy}.}
Regulations such as GDPR~\cite{GDPR2016} impose severe penalties for data leakage, granting users strict control over their data.
While traditional software focuses on database security, AI compliance also requires preventing the model \textit{itself} from leaking information.
\qarating{3}{5} highlights a unique AI risk: models inadvertently ``memorizing'' and regurgitating sensitive training data (e.g., financial records).
Even with non-sensitive data (e.g., browsing history), \qarating{8}{4} warns that compliance necessitates strict privacy measures to prevent re-identification attacks, ensuring the model respects the boundaries of data usage.

\begin{findingbox}
    \textbf{Perception 3}: Compliance acts as a legislative specification that translates legal mandates into prioritized quality attributes for AI models. 
    Regulations against harmful or illegal content require \textit{robustness}; auditability requirements mandate \textit{explainability}; anti-discrimination laws mandate \textit{fairness}; and data protection regimes prioritize \textit{privacy}, including safeguards against both direct data leakage and leakage through trained models.
\end{findingbox}

\subsubsection{\textbf{Explainability beyond compliance: perspectives for model developers and end-users}}

While the above perception highlights the necessity of regulatory compliance, the practical value of \textit{explainability} serves two distinct but complementary masters: the developers building the models and the end-users relying on them.

For \textbf{AI model developers}, \textit{explainability} serves as a critical diagnostic instrument.
It enables a deep inspection of internal mechanisms to identify weaknesses and guide targeted optimizations.
As described by \qarating{8}{4} and \qarating{2}{5}, this workflow involves isolating ``bad cases''—such as clearly irrelevant search results—and utilizing explainability tools to diagnose the root cause of these errors.
Crucially, the reliance on explainability as a debugging tool also underscores the high value they place on model \textit{maintainability}, as reflected in the high ratings (\qarating{8}{5} and \qarating{2}{5}).
While past surveys found developers used \textit{explainability} primarily for self-diagnosis~\cite{bhatt2020explainable}, our findings reveal a shift.
Today, practitioners emphasize a dual purpose: using explainability for debugging while also boosting end-user trust, aligning with evidence that transparency in AI assistants enhances trustworthiness~\cite{lyu2025my}.

For the \textbf{end-user}, \textit{explainability} supports ``black box'' predictions with trustable rationales.
In recommender systems, short, concrete explanations foster a sense of personal connection.
As \qarating{8}{4} illustrates, telling a customer `Recommend you this tea shop since you usually order boba around this time' creates a ``You got me!'' moment.
In technical domains like code vulnerability detection, this context supports critical decision-making.
\qarating{2}{5} notes that detailed explanations of why AI thinks a code is vulnerable help developers (as end-users) efficiently distinguish genuine vulnerabilities from false positives.
This necessity for functional explainability in vulnerability detection domain is further corroborated by recent academic literature (e.g.,~\cite{Chu_2024}).

\begin{findingbox}
    \textbf{Perception 4}: Beyond regulatory compliance, \textit{explainability} serves dual purposes in practice. 
    It remains a critical diagnostic tool for developers to improve model \textit{maintainability}, while also supporting end-user trust and human decision-making by making model outputs more interpretable and actionable.
\end{findingbox}

\subsection{RQ2: Challenges and Solutions in Quality Assurance}
\label{sec:rq2}

\textbf{Finding 1: To mitigate data imbalance arising from scarce positive cases, active learning proves effective for data acquisition, while traditional resampling and augmentation methods remain dominant in data synthesis due to the control they offer.}
Data imbalance acts as a silent killer for model quality, particularly in high-stakes domains involving rare events.
With positive cases often constituting less than 0.1\% of datasets (e.g., fraud detection in P3), the scarcity creates a ripple effect: it not only degrades correctness and robustness but also compromises fairness for protected minority groups (P8, P11) and creates unreliable, opaque interpretations that diminish explainability (P6).
To combat this, practitioners deploy a two-pronged strategy: smarter data acquisition and pragmatic data synthesis.

First, active learning serves as a precision tool for data acquisition.
Despite some academic skepticism regarding its utility~\cite{lowell2019practical}, our practitioners (P1, P3) champion its effectiveness with practical evidence.
Take P1's fake document detection as an example. After initial training on a small labeled dataset, a preliminary model is deployed to process real-time user uploads.
A selection policy (e.g., data points that exhibit high predictive uncertainty) then identifies samples for human verification and annotation.
The newly labeled data is then iteratively incorporated to retrain and redeploy the model, P1 reports that this process can obtain up to 10\% AUC improvement compared with random sampling.

Second, regarding data synthesis, industrial pragmatism diverges sharply from academic trends. 
While recent literature advocates for complex Generative AI (e.g., TabDDPM~\cite{kotelnikov2023tabddpm} for tabular data or DiffuseMix~\cite{Islam_2024_CVPR} for images), our practitioners (P1, P3, P6, P9) hesitate to adopt them due to quality instability, which is also questioned in previous surveys~\cite{nahar2023meta}.
Instead, they rely on `white-box' traditional methods.
For tabular data like fraud detection, established resampling techniques like SMOTE~\cite{Chawla_2002} remain the operational gold standard, reported by P3 and P6.
Similarly for imagery, P1 employs a robust augmentation pipeline that combines low-level perturbations (e.g., rotation, scaling, and JPEG recompression) with domain-specific tampering simulations.
In particular, it synthesizes realistic forgeries via copy-move and splicing operations, followed by post-processing steps such as boundary blending/smoothing and color-illumination harmonization to better match the surrounding context and mimic consistent acquisition artifacts.
As P1 summarizes, the ability to inject explicit domain knowledge often outweighs the theoretical allure of generative models: \textit{``My domain knowledge enables me to synthesize representative data using rule-based data augmentation.''}

\vspace*{0.1cm}
\noindent
\textbf{Finding 2: While data drift is a critical concern, monitoring complexities often compel the industry to bypass drift detection in favor of pragmatic retraining based on time or newly collected data volume.}
Data drift—the evolution of target distributions over time—is a known threat to model maintainability, it also threatens correctness and robustness, yet effectively monitoring it and determining the appropriate retraining trigger remains elusive.
As noted by Paleyes et al.~\cite{10.1145/3533378} and Nahar et al.~\cite{nahar2023meta}, current monitoring methods are often delayed and inaccurate.
Participants echoed this difficulty: P11 argued that retraining based on performance degradation is \textit{``too late,''} while P8 noted that direct monitoring target data distribution is also unreliable.
\textit{``Sometimes our metrics fail to capture data drift in time, while at other times they signal a drift to which our model is not sensitive.''}

Consequently, rather than relying on complex drift detection, P1, P6, P8 and P10 adopt two deterministic triggers.
The first is time-based scheduling tailored to business needs, such as P6's seasonal fraud detection updates or P8's weekly recommender refreshes.
The second trigger is data volume, where retraining is initiated by the accumulation of new labeled data—specifically, reaching a threshold for either the minority class count or the overall data magnitude.
For instance, P1 triggers retraining upon identifying 50 high-quality fake documents, whereas P10 relies on a total volume of 100 newly collected web pages.

\vspace*{0.1cm}
\noindent
\edit{
\textbf{Finding 3: To minimize operational costs while maintaining accuracy and latency at acceptable levels, practitioners jointly optimize hardware selection (CPU-based inference) and software techniques (model compression).}
Generally, achieving superior model performance necessitates higher operational costs and increased inference latency.
As discussed in \textbf{Perception 1}, while offline applications may compromise on latency to ensure accuracy at a lower cost, real-time applications operate under strict deadlines where every millisecond counts.
Furthermore, operational costs have become a critical constraint for sustainability.
To meet these dual demands of speed and thrift, we observe two industry practices often overlooked by previous surveys~\cite{nahar2023meta, you2025navigating}: the strategic pivot to CPU-based inference enabled by model compression.}

The economic motivation for this shift is particularly acute in Recommender Systems.
Unlike Large Language Models (LLMs) which are primarily compute-bound, Recommender Systems rely on massive Embedding Tables that can scale up to several Terabytes.
In this context, memory capacity is the bottleneck.
Scaling High Bandwidth Memory (HBM) on GPUs is exorbitantly expensive.
In stark contrast, equipping a standard CPU server with DDR5 memory is significantly cheaper.
P8 captures this trade-off succinctly: \textit{``a GPU with 80GB VRAM is even more expensive than 800GB RAM with CPU.''}
Consequently, transitioning to CPU-based deployment offers substantial cost savings, provided the latency and capability constraints can still be met.

This is where model compression becomes indispensable.
To ensure that CPU-based inference remains fast enough for real-time traffic and maintain the model capability, practitioners employ techniques such as pruning~\cite{salama2019pruning} to eliminate redundant parameters, quantization~\cite{gholami2022survey} to reduce numerical precision, and knowledge distillation~\cite{hinton2015distilling} to train compact student models using a larger teacher model's guidance.
As noted by P3, P6, and P8, this synergy allows them to leverage the cost-efficiency of high-memory CPUs without sacrificing user experience.

\vspace*{0.1cm}
\noindent
\textbf{Finding 4: AI-assisted annotation reduces labeling costs and can improve annotation quality.}
Manual annotation imposes substantial costs, reaching up to \$40 per article summary according to prior surveys~\cite{you2025navigating}.
This forces practitioners to continuously balance correctness against efficiency.
P6 adopts a conservative strategy involving two independent human annotators, effectively doubling annotation costs.
By contrast, P10 employs an AI-assisted workflow in which the model pre-labels data and a single human expert intervenes only to resolve disagreements.
While earlier work reports AI assistance primarily in test data annotation~\cite{you2025navigating}, we observe practitioners extending this approach to training data.
This shift reflects the growing need for frequent retraining, as discussed in Finding 2.
Notably, AI-assisted annotation can also enhance quality.
P10 reports that this hybrid workflow outperforms purely human annotation:
\textit{``We find the model accuracy even increases 5\% if AI is assisting the annotation compared with two humans. We observe that unlike two people who often share the same blind spots, the AI breaks that `false consensus', catching edge cases that humans would mutually overlook.''}

\vspace*{0.1cm}
\noindent
\textbf{Other Findings} Besides the key findings above, practitioners also mention many other challenges and mitigations that align with existing literature.
For example, consistent with You et al.~\cite{you2025navigating}, participants (P1, P8, P11) highlight the misalignment between offline accuracy and online business value. To mitigate this, they employ rapid canary releases on small user cohorts ($<5\%$). Models showing poor real-time business metrics trigger immediate rollbacks, while successful ones are gradually scaled to full deployment.
Similarly, participants echo the distinct need for active policing of Generative AI outputs.
While traditional ML prioritizes statistical accuracy, P3 and P14 emphasize that generative models demand strict \textit{output compliance} to prevent ``jailbreaks'' and toxicity. Mitigations include Reinforcement Learning from Human Feedback (RLHF) for internal alignment (P3) and external I/O guardrails like Llama Guard~\cite{inan2023llamaguardllmbasedinputoutput} (P14), corroborate recent findings~\cite{nahar2025beyond,you2025navigating}. In addition, we do not find that practitioners actively take measures to address model theft. For privacy-threatening attacks such as membership inference, no technical mitigation methods are adopted; instead, practitioners rely on organizational controls—for example, training highly sensitive models (e.g., P6’s fraud detection model) entirely in-house and deploying them in isolated environments with no external access.
The challenge of selecting explainability methods doesn't have a generalizable best practice. Practitioners (P3, P6, P8) characterize this process as navigating trade-offs rather than applying a standard procedure, aligning with prior work~\cite{clement2023xair}. As P6 observes, choosing an appropriate explanation method remains largely a trial-and-error process.

\subsection{RQ3: To What Extent Are Our Findings Recognized in Practice?}
\label{sec:rq3}

To assess how our findings resonate with industry practice, we conducted a practitioner survey following the methodology in Section~\ref{subsec:survey}.
In addition to Likert-scale ratings, respondents were invited to provide optional free-text comments explaining their agreement or disagreement.
This section reports both quantitative agreement levels and qualitative reflections that clarify the sources of consensus as well as remaining reservations.

Participants rate each finding on a 5-point Likert scale ranging from \textit{strongly disagree} (1) to \textit{strongly agree} (5), with an additional option for \textit{I don't know}.
Building on prior work~\cite{8804445}, we categorize findings into three levels of acknowledgment: Well-Acknowledged (average $\geq 4.0$), Marginally Agreeable ($3.5 \leq$ average $< 4.0$), and Controversial/Not Acknowledged (average $< 3.5$).
For findings composed of multiple components (Perception~3 and Challenge \& Solution~1), each component is evaluated independently.
Since all components are well acknowledged, we report their aggregated means for conciseness.
Table~\ref{tab:takeaways} summarizes the results: seven findings are well acknowledged and one is marginally agreeable.
No findings fall into the controversial category within our sample, suggesting broad recognition across the surveyed teams and domains.
Full survey results are available in the replication package.

\vspace*{0.1cm}
\noindent
\textbf{Perception 1: Efficiency is a dominant constraint on correctness in real-time AI.}
This perception receives an average score of \textbf{4.16}, indicating strong endorsement.
Practitioners consistently emphasize that in real-time applications—such as fraud detection and live recommendation—strict latency and cost budgets frequently necessitate accepting modest performance degradations.
As one respondent explains, \textit{``in most real-time use cases like fraud detection or live recommendation systems, this trade-off holds true.''}
At the same time, a minority caution against overly aggressive efficiency constraints.
One practitioner notes that \textit{``insufficient capability causes struggles with complex tasks,''}
highlighting the importance of carefully selecting acceptable correctness thresholds.

\vspace*{0.1cm}
\noindent
\textbf{Perception 2: Deployability and scalability burdens are largely shifted away from AI developers via platform-based microservices.}
This perception attains an average score of \textbf{4.02}, reflecting the widespread adoption of containerization and cloud platforms.
Many respondents highlight that modern infrastructure tooling substantially reduces deployment friction.
One practitioner remarks, \textit{``cannot agree more, thanks to Docker and Kubernetes.''}
Nevertheless, some reservations emerge.
A small number of respondents point out that service decoupling can introduce new risks, cautioning that \textit{``you should also consider potential vulnerability issues''} and that \textit{``it is still hard to orchestrate and monitor models''} in complex deployments.

\vspace*{0.1cm}
\noindent
\textbf{Perception 3: Compliance selectively elevates the importance of robustness, explainability, fairness, and privacy.}
Each quality attribute is evaluated independently and is well acknowledged (robustness: \textbf{4.02}, explainability: \textbf{4.30}, fairness: \textbf{4.24}, privacy: \textbf{4.38}).
Participants raise no substantive objections to this perception.
Instead, they emphasize that regulatory requirements render these attributes non-negotiable in many domains.
For example, one respondent observes that \textit{``in generative AI, robustness is no longer just a technical metric but a legal necessity,''}
while another notes that \textit{``any model using demographic data is heavily scrutinized, making fairness non-negotiable.''}
Privacy obligations are likewise underscored, with practitioners stressing the need to comply with regulations such as GDPR and CCPA.

\vspace*{0.1cm}
\noindent
\textbf{Perception 4: Explainability is essential for user trust and model debugging.}
This perception receives an average score of \textbf{4.18}.
Respondents highlight the dual role of explainability in supporting both developer-side debugging and end-user trust.
One practitioner succinctly captures this view, stating that \textit{``explainability isn't just about meeting rules.''}
At the same time, a minority caution against overreliance on post-hoc explanations.
As one respondent warns, \textit{``post-hoc explainability is not always helpful if the underlying model is fundamentally flawed,''}
emphasizing that core model quality must be addressed first.

\vspace*{0.1cm}
\noindent
\textbf{Challenge \& Solution 1: Active learning optimizes data representativeness, while resampling and augmentation support data synthesis.}
This finding is well acknowledged with an average score of \textbf{4.13}.
Practitioners broadly agree that active learning enables more efficient use of limited labeling resources by prioritizing uncertain or minority-class instances.
One respondent notes that \textit{``it prioritizes labelling uncertain or minority.''}
However, some also point to practical limitations.
As one practitioner cautions, \textit{``its main limitation lies in the added operational complexity and reliance on timely human annotation.''}
Traditional resampling and augmentation methods are likewise strongly endorsed (average score: \textbf{4.14}), with respondents valuing their controllability.
One respondent notes, \textit{``it lets me control the exact oversampling ratios for minority classes.''}

\vspace*{0.1cm}
\noindent
\textbf{Challenge \& Solution 2: Complex data drift monitoring is often bypassed in favor of pragmatic retraining.}
This finding receives an average score of \textbf{4.00}.
Most practitioners report replacing complex real-time drift detection with scheduled retraining or retraining triggered by the accumulation of new data.
One respondent summarizes this practice by noting that \textit{``many teams rely on scheduled retraining because real-time drift detection is costly and complex.''}
At the same time, some express caution.
As one practitioner observes, \textit{``new data does not necessarily mean better models and might just mean wasted time and computation resource,''}
underscoring the need for quality control alongside retraining.

\vspace*{0.1cm}
\noindent
\textbf{Challenge \& Solution 3: Model compression paired with CPU-based inference optimizes cost and latency.}
With an average score of \textbf{4.10}, this strategy is widely recognized, particularly in cost-sensitive deployments such as customer-facing chatbots.
Respondents emphasize that techniques such as quantization enable efficient CPU-based inference and help avoid the higher costs associated with GPUs.
One practitioner describes this approach as \textit{``a standard practice''} for customer support systems.
However, respondents also note clear boundary conditions.
For large models with high-throughput or reasoning-intensive requirements, GPU-based inference remains necessary.
As one practitioner explains, \textit{``for large-scale models (70B+ for ours) with high-throughput requirements, GPU-based inference is still a must.''}

\vspace*{0.1cm}
\noindent
\textbf{Challenge \& Solution 4: AI-assisted annotation improves efficiency while maintaining acceptable quality.}
This finding is classified as marginally agreeable with an average score of \textbf{3.98}, based on our predefined threshold.
Practitioners generally acknowledge that AI-assisted annotation can substantially reduce labeling costs.
One respondent reports that \textit{``AI-assisted data annotation reduces costs by 30--50\%.''} 
At the same time, respondents in highly specialized domains emphasize the continued importance of human expertise.
As one practitioner cautions, \textit{``rare disease diagnosis requires extensive expertise,''}
highlighting the domain-dependent applicability of AI-assisted annotation workflows.

\section{Discussion}
\label{sec:discussion}

\subsection{Suggestions to Practitioners}
\label{sec:discussion-suggestions}

Practitioners are advised to adopt a unified set of practices that balance technical constraints with industry evolution. To ensure \textit{correctness} and \textit{robustness}, one should address class imbalance through active learning, while retaining traditional resampling and augmentation for their precise distributional control. To maximize \textit{efficiency} in real-time settings, we recommend navigating the cost-latency-capability tradeoff by coordinating hardware selection with model compression, and leveraging AI-assisted labeling to reduce costs while improving quality. Regarding \textit{maintainability}, strategies should favor pragmatism by replacing complex drift detection with regular retraining pipelines, while \textit{deployability} and \textit{scalability} are best managed through containerized microservices and dynamic platform scaling. Finally, compliance must be translated directly into attributes—ensuring \textit{robustness} for safety, \textit{fairness} for anti-discrimination, and \textit{privacy} for data protection—while \textit{explainability} should be expanded beyond debugging to foster user trust and meet auditability standards.

\subsection{\textbf{Threats to Validity}}
\label{sec:discussion-threats}
\noindent\textbf{Internal.}
Subjectivity in data interpretation and instrument design poses a primary threat.
To mitigate bias in our study design, as detailed in Section~\ref{subsec:attributes_selection}, the selection of key quality attributes is rigorously validated through literature comparison, pilot interviews with two practitioners, and questionnaire.
For data analysis, two authors independently double-code all transcripts and reconcile differences.
We further ensure data accuracy by conducting member-checks with five participants.
The codebook is provided in the replication package to ensure procedural transparency.

\noindent\textbf{External.}
A key threat is the limited generalizability of findings derived from 15 interviews.
We mitigate this by ensuring diverse roles, focus areas, countries, and company sizes for our interviewees.
Crucially, we triangulate our qualitative results through a validation survey with 50 AI practitioners.
The survey results confirm that our findings are widely acknowledged by the broader community, with the exception of ``AI-assisted data annotation,'' which is marginally agreeable.
This triangulation significantly strengthens the transferability of our insights beyond interviewees.

\section{Related Work}
\label{sec:related_work}
\vspace{-0.2em}

\subsection{Challenges in Industrial AI Development}
\label{sec:challenges_in_industrial_ai_development}
\vspace{-0.2em}
Given the unique nature of AI models and the research-industry gap, many studies have explored challenges in industrial AI development.
Hill et al.~\cite{hill2016trials} interview 11 practitioners, while Zhang et al.~\cite{zhang2019software} interview 8 and survey 195 more, revealing that AI's data-driven nature brings new challenges throughout the development lifecycle.
Unlike the above multi-domain studies, Amershi et al.~\cite{8804457} present an in-depth case study of Microsoft's AI development, revealing organization-specific challenges and solutions—such as the greater complexity in data discovery, management, and versioning—that distinguish AI from traditional software engineering.
Arpteg et al.~\cite{arpteg2018software} examine software engineering challenges in AI development across seven industrial projects, their study focuses on inspiring future research, thus not providing solutions.

Besides interviewing practitioners, challenges can also be identified from the literature.
For example, Paleyes et al.~\cite{10.1145/3533378}, Nahar et al.~\cite{nahar2023meta}, and Lwakatare et al.~\cite{lwakatare2020large} mine the literature and identify challenges faced in each stage of ML deployment workflow.
Paleyes et al.~\cite{10.1145/3533378} note that many ML challenges mirror those in fields such as software engineering, human-computer interaction, and policymaking, suggesting ML can benefit from their established knowledge.
Nahar et al.~\cite{nahar2023meta} notice that there is more consensus on what is a challenge than on how to solve it.
Lwakatare et al.~\cite{lwakatare2020large} emphasize that machine learning brings new requirements to both software development process and the practices of organizations.
These interview studies and literature reviews have offered valuable landscape-level insights—for example, noting that \textit{`feature engineering is difficult'} or that \textit{`evaluation metrics can be too narrow'}.
Furthermore, literature reviews on industry challenges may not fully reflect current practices, as both studies include many pre-2020 papers, limiting relevance to today's rapidly evolving AI landscape.

Sinha et al.~\cite{sinha2024challenges} draw on literature, case studies, industry reports, and public datasets to surface AI development and deployment challenges.
While several overlaps with ours, their study leaves most challenges unaddressed. Our work confirms these issues and provides practitioner-validated remedies.

Besides challenges, several studies~\cite{8804457, 10.1145/3382494.3410681, chouliaras2023best, serban2024software} have identified best practices for AI model development, these practices are rather general and not specific to challenges.

Several studies examine the testing phase of AI development, as it directly reflects its quality.
Song et al.~\cite{song2022exploring} review papers on ML testing, extracting challenges and potential solutions.
They find that none of five academic papers relevant to their partner company, Axis Communications, address their real-world AI testing needs—highlighting the research-industry gap.
Zhang et al.~\cite{zhang2019machine} review ML testing techniques in academia literature for various quality attributes.
Borg et al.~\cite{borg2020safely} investigate verification and validation challenges for ML-powered autonomous driving systems, finding that ISO 26262 (vehicle safety standard) conflicts with deep neural network characteristics.
They suggest adopting aerospace practices and systems-based safety approaches for automotive applications.
You et al.~\cite{you2025navigating} explore challenges and solutions for AI testing in industry through practitioner interviews.
Unlike their work, our study goes beyond testing to cover all processes impacting AI quality, discussing challenges like generative model compliance and offering industry-validated solutions such as human-in-the-loop and role-based filtering.

\vspace{-0.2em}
\subsection{AI Quality Attributes}
\label{sec:ai_quality_attributes}
\vspace{-0.2em}
A comprehensive quality model is essential to evaluate AI software beyond correctness. Traditional software quality models, such as McCall et al.~\cite{general1977factors} and ISO/IEC 9126~\cite{organization1991iso} (later ISO/IEC 25010~\cite{iso25010-2011}), lay the groundwork but lack AI-specific focus.
Later, ISO/IEC 25059 \cite{iso25059-2023} introduces AI-oriented quality attributes.
Meanwhile, researchers continue to examine AI-related non-functional requirements (NFRs) and domain-specific models.
For example, Habibullah et al.~\cite{habibullah2022non} cluster NFRs for ML systems, Siebert et al.~\cite{siebert2022construction} propose tailored industrial models, and Villamizar et al.~\cite{villamizar2024identifying} identify ML quality attributes through a perspective-based approach.

There are several studies that identify challenges under the taxonomy of NFRs.
De Martino et al.~\cite{de2025classification} identify 30 NFRs and over 23 challenges requiring further research from ML literature.
They focus on academic literature, while we focus on industry practice.
Their challenges are broad aiming to inspire future research, while we present concrete, attribute-specific challenges with practitioner-validated solutions to directly aid practitioners.
Horkoff~\cite{8920538} outline the challenges and research directions for NFRs.
Building on this foundation, Habibullah et al.~\cite{habibullah2023non} analyze NFRs for ML systems by 10 interviews.
Unlike their focus on overall NFR importance, we analyze how and why quality attribute priorities differ by domain.
While they identify generic NFR challenges such as \textit{safety is difficult to guarantee}~\cite{habibullah2023non} and the \textit{understanding of NFRs for AI is fragmented and incomplete}~\cite{8920538}, we target attribute-specific challenges, such as data drift undermining robustness~\cite{lyu2023chronos}.

\vspace{-0.2em}
\section{Conclusion and Future Work}
\label{sec:conclusion}
\vspace{-0.2em}
This study focuses on the perceptions, challenges, and solutions of different quality attributes of AI models.
Nine key quality attributes are identified from the literature that are relevant to industry AI model development.
We conduct an interview study with fifteen AI industry experts from various roles, countries, and company sizes.
Four perceptions and four challenges along with the corresponding solutions are identified.
These eight findings are further quantitatively validated by a survey of fifty industry practitioners, and seven of the findings are well-acknowledged.
Our findings can guide researchers to focus on attributes valued by practitioners when designing new techniques, while avoiding approaches that improve one attribute at the expense of others deemed more critical.
Peer-validated solutions can also serve as tools for practitioners to address similar challenges.
Future research should expand beyond AI models to examine other critical components of AI systems.

\textbf{Data availability: replication package at \url{https://github.com/JamesNolan17/AIModelQuality}.}

\section{Acknowledgments}
This research is supported by the Ministry of Education, Singapore under its Academic Research Fund Tier 3 (Award ID: MOET32020-0004). Any opinions, findings and conclusions or recommendations expressed in this material are those of the author(s) and do not reflect the views of the Ministry of Education, Singapore.

\balance{}
\bibliographystyle{ACM-Reference-Format}
\bibliography{reference}


\begin{thebibliography}{64}


\ifx \showCODEN    \undefined \def \showCODEN     #1{\unskip}     \fi
\ifx \showDOI      \undefined \def \showDOI       #1{#1}\fi
\ifx \showISBNx    \undefined \def \showISBNx     #1{\unskip}     \fi
\ifx \showISBNxiii \undefined \def \showISBNxiii  #1{\unskip}     \fi
\ifx \showISSN     \undefined \def \showISSN      #1{\unskip}     \fi
\ifx \showLCCN     \undefined \def \showLCCN      #1{\unskip}     \fi
\ifx \shownote     \undefined \def \shownote      #1{#1}          \fi
\ifx \showarticletitle \undefined \def \showarticletitle #1{#1}   \fi
\ifx \showURL      \undefined \def \showURL       {\relax}        \fi
\providecommand\bibfield[2]{#2}
\providecommand\bibinfo[2]{#2}
\providecommand\natexlab[1]{#1}
\providecommand\showeprint[2][]{arXiv:#2}

\bibitem[Pre({[n.\,d.]})]%
        {Press_2022}
 \bibinfo{year}{[n.\,d.]}\natexlab{}.
\newblock \bibinfo{title}{Nearly 400 car crashes in 11 months involved
  Automated Tech}.
\newblock
\newblock
\urldef\tempurl%
\url{https://tinyurl.com/22ux4458}
\showURL{%
\tempurl}


\bibitem[tri({[n.\,d.]})]%
        {triton}
 \bibinfo{year}{[n.\,d.]}\natexlab{}.
\newblock \bibinfo{title}{NVIDIA Triton Inference Server}.
\newblock
\newblock
\urldef\tempurl%
\url{https://developer.nvidia.com/nvidia-triton-inference-server}
\showURL{%
\tempurl}


\bibitem[iso(2011)]%
        {iso25010-2011}
 \bibinfo{year}{2011}\natexlab{}.
\newblock \bibinfo{title}{{ISO/IEC 25010: Systems and software engineering ---
  Systems and software Quality Requirements and Evaluation (SQuaRE) --- System
  and software quality models}}.
\newblock
\newblock


\bibitem[iso(2023)]%
        {iso25059-2023}
 \bibinfo{year}{2023}\natexlab{}.
\newblock \bibinfo{title}{ISO/IEC 25059:2023: Software engineering --- Systems
  and software Quality Requirements and Evaluation (SQuaRE) --- Quality model
  for AI systems}.
\newblock
\newblock


\bibitem[eua(2024)]%
        {euaiact2024}
 \bibinfo{year}{2024}\natexlab{}.
\newblock \bibinfo{title}{{Regulation (EU) 2024/1689} (Artificial Intelligence
  Act)}.
\newblock
  \bibinfo{howpublished}{\url{http://data.europa.eu/eli/reg/2024/1689/oj}}.
\newblock
\newblock
\shownote{OJ L 2024/1689, 12 July 2024}.


\bibitem[eu_(2025)]%
        {eu_aml}
 \bibinfo{year}{2025}\natexlab{}.
\newblock \bibinfo{title}{Anti-money laundering and countering the financing of
  terrorism at EU level}.
\newblock
  \bibinfo{howpublished}{\url{https://finance.ec.europa.eu/financial-crime/anti-money-laundering-and-countering-financing-terrorism-eu-level_en}}.
\newblock


\bibitem[Amershi et~al\mbox{.}(2019)]%
        {8804457}
\bibfield{author}{\bibinfo{person}{Saleema Amershi}, \bibinfo{person}{Andrew
  Begel}, \bibinfo{person}{Christian Bird}, \bibinfo{person}{Robert DeLine},
  \bibinfo{person}{Harald Gall}, \bibinfo{person}{Ece Kamar},
  \bibinfo{person}{Nachiappan Nagappan}, \bibinfo{person}{Besmira Nushi}, {and}
  \bibinfo{person}{Thomas Zimmermann}.} \bibinfo{year}{2019}\natexlab{}.
\newblock \showarticletitle{Software Engineering for Machine Learning: A Case
  Study}. In \bibinfo{booktitle}{\emph{2019 IEEE/ACM 41st International
  Conference on Software Engineering: Software Engineering in Practice
  (ICSE-SEIP)}}. \bibinfo{pages}{291--300}.
\newblock
\urldef\tempurl%
\url{https://doi.org/10.1109/ICSE-SEIP.2019.00042}
\showDOI{\tempurl}


\bibitem[Arpteg et~al\mbox{.}(2018)]%
        {arpteg2018software}
\bibfield{author}{\bibinfo{person}{Anders Arpteg}, \bibinfo{person}{Bj{\"o}rn
  Brinne}, \bibinfo{person}{Luka Crnkovic-Friis}, {and} \bibinfo{person}{Jan
  Bosch}.} \bibinfo{year}{2018}\natexlab{}.
\newblock \showarticletitle{Software engineering challenges of deep learning}.
  In \bibinfo{booktitle}{\emph{2018 44th euromicro conference on software
  engineering and advanced applications (SEAA)}}. IEEE,
  \bibinfo{pages}{50--59}.
\newblock


\bibitem[Baldwin et~al\mbox{.}(2025)]%
        {baldwin2025ethicspracticesaidevelopment}
\bibfield{author}{\bibinfo{person}{Wilder Baldwin}, \bibinfo{person}{Sepideh
  Ghanavati}, {and} \bibinfo{person}{Manuel Woersdoerfer}.}
  \bibinfo{year}{2025}\natexlab{}.
\newblock \bibinfo{title}{Ethics Practices in AI Development: An Empirical
  Study Across Roles and Regions}.
\newblock
\newblock
\showeprint[arxiv]{2508.09219}~[cs.CY]
\urldef\tempurl%
\url{https://arxiv.org/abs/2508.09219}
\showURL{%
\tempurl}


\bibitem[Bao et~al\mbox{.}(2022)]%
        {bao2022artificial}
\bibfield{author}{\bibinfo{person}{Yang Bao}, \bibinfo{person}{Gilles Hilary},
  {and} \bibinfo{person}{Bin Ke}.} \bibinfo{year}{2022}\natexlab{}.
\newblock \showarticletitle{Artificial intelligence and fraud detection}.
\newblock \bibinfo{journal}{\emph{Innovative Technology at the Interface of
  Finance and Operations: Volume I}} (\bibinfo{year}{2022}),
  \bibinfo{pages}{223--247}.
\newblock


\bibitem[Barmer et~al\mbox{.}(2021)]%
        {Barmer2021}
\bibfield{author}{\bibinfo{person}{Hollen Barmer}, \bibinfo{person}{Rachel
  Dzombak}, \bibinfo{person}{Matthew Gaston}, \bibinfo{person}{Vijaykumar
  Palat}, \bibinfo{person}{Frank Redner}, \bibinfo{person}{Tanisha Smith},
  {and} \bibinfo{person}{John Wohlbier}.} \bibinfo{year}{2021}\natexlab{}.
\newblock \showarticletitle{{Scalable AI}}.
\newblock  (\bibinfo{date}{9} \bibinfo{year}{2021}).
\newblock
\urldef\tempurl%
\url{https://doi.org/10.1184/R1/16560273.v1}
\showDOI{\tempurl}


\bibitem[Bhatt et~al\mbox{.}(2020)]%
        {bhatt2020explainable}
\bibfield{author}{\bibinfo{person}{Umang Bhatt}, \bibinfo{person}{Alice Xiang},
  \bibinfo{person}{Shubham Sharma}, \bibinfo{person}{Adrian Weller},
  \bibinfo{person}{Ankur Taly}, \bibinfo{person}{Yunhan Jia},
  \bibinfo{person}{Joydeep Ghosh}, \bibinfo{person}{Ruchir Puri},
  \bibinfo{person}{Jos{\'e}~MF Moura}, {and} \bibinfo{person}{Peter
  Eckersley}.} \bibinfo{year}{2020}\natexlab{}.
\newblock \showarticletitle{Explainable machine learning in deployment}. In
  \bibinfo{booktitle}{\emph{Proceedings of the 2020 conference on fairness,
  accountability, and transparency}}. \bibinfo{pages}{648--657}.
\newblock


\bibitem[Boden(2016)]%
        {boden2016ai}
\bibfield{author}{\bibinfo{person}{Margaret~A Boden}.}
  \bibinfo{year}{2016}\natexlab{}.
\newblock \bibinfo{booktitle}{\emph{AI: Its nature and future}}.
\newblock \bibinfo{publisher}{Oxford University Press}.
\newblock


\bibitem[Borg et~al\mbox{.}(2020)]%
        {borg2020safely}
\bibfield{author}{\bibinfo{person}{Markus Borg}, \bibinfo{person}{Cristofer
  Englund}, \bibinfo{person}{Krzysztof Wnuk}, \bibinfo{person}{Boris Duran},
  \bibinfo{person}{Christoffer Levandowski}, \bibinfo{person}{Shenjian Gao},
  \bibinfo{person}{Yanwen Tan}, \bibinfo{person}{Henrik Kaijser},
  \bibinfo{person}{Henrik L{\"o}nn}, {and} \bibinfo{person}{Jonas
  T{\"o}rnqvist}.} \bibinfo{year}{2020}\natexlab{}.
\newblock \showarticletitle{Safely entering the deep: A review of verification
  and validation for machine learning and a challenge elicitation in the
  automotive industry}.
\newblock \bibinfo{journal}{\emph{Journal of Automotive Software Engineering}}
  \bibinfo{volume}{1}, \bibinfo{number}{1} (\bibinfo{year}{2020}),
  \bibinfo{pages}{1--19}.
\newblock


\bibitem[Chawla et~al\mbox{.}(2002)]%
        {Chawla_2002}
\bibfield{author}{\bibinfo{person}{N.~V. Chawla}, \bibinfo{person}{K.~W.
  Bowyer}, \bibinfo{person}{L.~O. Hall}, {and} \bibinfo{person}{W.~P.
  Kegelmeyer}.} \bibinfo{year}{2002}\natexlab{}.
\newblock \showarticletitle{{SMOTE}: Synthetic Minority Over-sampling
  Technique}.
\newblock \bibinfo{journal}{\emph{Journal of Artificial Intelligence Research}}
   \bibinfo{volume}{16} (\bibinfo{date}{jun} \bibinfo{year}{2002}),
  \bibinfo{pages}{321--357}.
\newblock
\urldef\tempurl%
\url{https://doi.org/10.1613/jair.953}
\showDOI{\tempurl}


\bibitem[Chen et~al\mbox{.}(2025)]%
        {chen2025secureagentbench}
\bibfield{author}{\bibinfo{person}{Junkai Chen}, \bibinfo{person}{Huihui
  Huang}, \bibinfo{person}{Yunbo Lyu}, \bibinfo{person}{Junwen An},
  \bibinfo{person}{Jieke Shi}, \bibinfo{person}{Chengran Yang},
  \bibinfo{person}{Ting Zhang}, \bibinfo{person}{Haoye Tian},
  \bibinfo{person}{Yikun Li}, \bibinfo{person}{Zhenhao Li}, {et~al\mbox{.}}}
  \bibinfo{year}{2025}\natexlab{}.
\newblock \showarticletitle{SecureAgentBench: Benchmarking Secure Code
  Generation under Realistic Vulnerability Scenarios}.
\newblock \bibinfo{journal}{\emph{arXiv preprint arXiv:2509.22097}}
  (\bibinfo{year}{2025}).
\newblock


\bibitem[Chouliaras et~al\mbox{.}(2023)]%
        {chouliaras2023best}
\bibfield{author}{\bibinfo{person}{Georgios~Christos Chouliaras},
  \bibinfo{person}{Kornel Kie{\l}czewski}, \bibinfo{person}{Amit Beka},
  \bibinfo{person}{David Konopnicki}, {and} \bibinfo{person}{Lucas Bernardi}.}
  \bibinfo{year}{2023}\natexlab{}.
\newblock \showarticletitle{Best Practices for Machine Learning Systems: An
  Industrial Framework for Analysis and Optimization}.
\newblock \bibinfo{journal}{\emph{arXiv preprint arXiv:2306.13662}}
  (\bibinfo{year}{2023}).
\newblock


\bibitem[Chu et~al\mbox{.}(2024)]%
        {Chu_2024}
\bibfield{author}{\bibinfo{person}{Zhaoyang Chu}, \bibinfo{person}{Yao Wan},
  \bibinfo{person}{Qian Li}, \bibinfo{person}{Yang Wu}, \bibinfo{person}{Hongyu
  Zhang}, \bibinfo{person}{Yulei Sui}, \bibinfo{person}{Guandong Xu}, {and}
  \bibinfo{person}{Hai Jin}.} \bibinfo{year}{2024}\natexlab{}.
\newblock \showarticletitle{Graph Neural Networks for Vulnerability Detection:
  A Counterfactual Explanation}. In \bibinfo{booktitle}{\emph{Proceedings of
  the 33rd ACM SIGSOFT International Symposium on Software Testing and
  Analysis}} \emph{(\bibinfo{series}{ISSTA ’24})}. \bibinfo{publisher}{ACM},
  \bibinfo{pages}{389–401}.
\newblock
\urldef\tempurl%
\url{https://doi.org/10.1145/3650212.3652136}
\showDOI{\tempurl}


\bibitem[Clement et~al\mbox{.}(2023)]%
        {clement2023xair}
\bibfield{author}{\bibinfo{person}{Tobias Clement}, \bibinfo{person}{Nils
  Kemmerzell}, \bibinfo{person}{Mohamed Abdelaal}, {and}
  \bibinfo{person}{Michael Amberg}.} \bibinfo{year}{2023}\natexlab{}.
\newblock \showarticletitle{XAIR: a systematic metareview of explainable AI
  (XAI) aligned to the software development process}.
\newblock \bibinfo{journal}{\emph{Machine Learning and Knowledge Extraction}}
  \bibinfo{volume}{5}, \bibinfo{number}{1} (\bibinfo{year}{2023}),
  \bibinfo{pages}{78--108}.
\newblock


\bibitem[Company et~al\mbox{.}(1977)]%
        {general1977factors}
\bibfield{author}{\bibinfo{person}{General~Electric Company},
  \bibinfo{person}{J.A. McCall}, \bibinfo{person}{P.K. Richards},
  \bibinfo{person}{G.F. Walters}, \bibinfo{person}{Rome Air~Development
  Center}, {and} \bibinfo{person}{United States. Air Force. Systems Command.
  Electronic~Systems Division}.} \bibinfo{year}{1977}\natexlab{}.
\newblock \bibinfo{booktitle}{\emph{Factors in Software Quality}}.
\newblock \bibinfo{publisher}{Information Systems Programs, General Electric
  Company}.
\newblock


\bibitem[Cruzes and Dyba(2011)]%
        {10.1109/ESEM.2011.36}
\bibfield{author}{\bibinfo{person}{Daniela~S. Cruzes} {and}
  \bibinfo{person}{Tore Dyba}.} \bibinfo{year}{2011}\natexlab{}.
\newblock \showarticletitle{Recommended Steps for Thematic Synthesis in
  Software Engineering}. In \bibinfo{booktitle}{\emph{Proceedings of the 2011
  International Symposium on Empirical Software Engineering and Measurement}}
  \emph{(\bibinfo{series}{ESEM '11})}. \bibinfo{publisher}{IEEE Computer
  Society}, \bibinfo{address}{USA}, \bibinfo{pages}{275–284}.
\newblock
\showISBNx{9780769546049}
\urldef\tempurl%
\url{https://doi.org/10.1109/ESEM.2011.36}
\showDOI{\tempurl}


\bibitem[De~Martino and Palomba(2025)]%
        {de2025classification}
\bibfield{author}{\bibinfo{person}{Vincenzo De~Martino} {and}
  \bibinfo{person}{Fabio Palomba}.} \bibinfo{year}{2025}\natexlab{}.
\newblock \showarticletitle{Classification and challenges of non-functional
  requirements in ML-enabled systems: A systematic literature review}.
\newblock \bibinfo{journal}{\emph{Information and Software Technology}}
  (\bibinfo{year}{2025}), \bibinfo{pages}{107678}.
\newblock


\bibitem[{European Union}(2016)]%
        {GDPR2016}
\bibfield{author}{\bibinfo{person}{{European Union}}.}
  \bibinfo{year}{2016}\natexlab{}.
\newblock \showarticletitle{Regulation (EU) 2016/679 of the European Parliament
  and of the Council of 27 April 2016 on the protection of natural persons with
  regard to the processing of personal data and on the free movement of such
  data, and repealing Directive 95/46/EC (General Data Protection Regulation)}.
\newblock \bibinfo{journal}{\emph{Official Journal of the European Union}}
  \bibinfo{volume}{L119} (\bibinfo{date}{May} \bibinfo{year}{2016}),
  \bibinfo{pages}{1--88}.
\newblock
\urldef\tempurl%
\url{https://eur-lex.europa.eu/eli/reg/2016/679/oj/eng}
\showURL{%
\tempurl}
\newblock
\shownote{Text with EEA relevance}.


\bibitem[Gholami et~al\mbox{.}(2022)]%
        {gholami2022survey}
\bibfield{author}{\bibinfo{person}{Amir Gholami}, \bibinfo{person}{Sehoon Kim},
  \bibinfo{person}{Zhen Dong}, \bibinfo{person}{Zhewei Yao},
  \bibinfo{person}{Michael~W Mahoney}, {and} \bibinfo{person}{Kurt Keutzer}.}
  \bibinfo{year}{2022}\natexlab{}.
\newblock \showarticletitle{A survey of quantization methods for efficient
  neural network inference}.
\newblock In \bibinfo{booktitle}{\emph{Low-power computer vision}}.
  \bibinfo{publisher}{Chapman and Hall/CRC}, \bibinfo{pages}{291--326}.
\newblock


\bibitem[Gong et~al\mbox{.}(2022)]%
        {acsac2022gong}
\bibfield{author}{\bibinfo{person}{Chen Gong}, \bibinfo{person}{Zhou Yang},
  \bibinfo{person}{Yunpeng Bai}, \bibinfo{person}{Jieke Shi},
  \bibinfo{person}{Arunesh Sinha}, \bibinfo{person}{Bowen Xu},
  \bibinfo{person}{David Lo}, \bibinfo{person}{Xinwen Hou}, {and}
  \bibinfo{person}{Guoliang Fan}.} \bibinfo{year}{2022}\natexlab{}.
\newblock \showarticletitle{Curiosity-Driven and Victim-Aware Adversarial
  Policies}. In \bibinfo{booktitle}{\emph{Proceedings of the 38th Annual
  Computer Security Applications Conference}} (Austin, TX, USA)
  \emph{(\bibinfo{series}{ACSAC '22})}. \bibinfo{publisher}{Association for
  Computing Machinery}, \bibinfo{address}{New York, NY, USA},
  \bibinfo{pages}{186–200}.
\newblock
\showISBNx{9781450397599}
\urldef\tempurl%
\url{https://doi.org/10.1145/3564625.3564636}
\showDOI{\tempurl}


\bibitem[Goodfellow et~al\mbox{.}(2014)]%
        {goodfellow2014generative}
\bibfield{author}{\bibinfo{person}{Ian~J. Goodfellow}, \bibinfo{person}{Jean
  Pouget-Abadie}, \bibinfo{person}{Mehdi Mirza}, \bibinfo{person}{Bing Xu},
  \bibinfo{person}{David Warde-Farley}, \bibinfo{person}{Sherjil Ozair},
  \bibinfo{person}{Aaron Courville}, {and} \bibinfo{person}{Yoshua Bengio}.}
  \bibinfo{year}{2014}\natexlab{}.
\newblock \bibinfo{title}{Generative Adversarial Networks}.
\newblock
\newblock
\showeprint[arxiv]{1406.2661}~[stat.ML]


\bibitem[Goodman(1961)]%
        {goodman1961snowball}
\bibfield{author}{\bibinfo{person}{Leo~A Goodman}.}
  \bibinfo{year}{1961}\natexlab{}.
\newblock \showarticletitle{Snowball sampling}.
\newblock \bibinfo{journal}{\emph{The annals of mathematical statistics}}
  (\bibinfo{year}{1961}), \bibinfo{pages}{148--170}.
\newblock


\bibitem[Habibullah et~al\mbox{.}(2022)]%
        {habibullah2022non}
\bibfield{author}{\bibinfo{person}{Khan~Mohammad Habibullah},
  \bibinfo{person}{Gregory Gay}, {and} \bibinfo{person}{Jennifer Horkoff}.}
  \bibinfo{year}{2022}\natexlab{}.
\newblock \showarticletitle{Non-functional requirements for machine learning:
  An exploration of system scope and interest}. In
  \bibinfo{booktitle}{\emph{Proceedings of the 1st Workshop on Software
  Engineering for Responsible AI}}. \bibinfo{pages}{29--36}.
\newblock


\bibitem[Habibullah et~al\mbox{.}(2023)]%
        {habibullah2023non}
\bibfield{author}{\bibinfo{person}{Khan~Mohammad Habibullah},
  \bibinfo{person}{Gregory Gay}, {and} \bibinfo{person}{Jennifer Horkoff}.}
  \bibinfo{year}{2023}\natexlab{}.
\newblock \showarticletitle{Non-functional requirements for machine learning:
  Understanding current use and challenges among practitioners}.
\newblock \bibinfo{journal}{\emph{Requirements Engineering}}
  \bibinfo{volume}{28}, \bibinfo{number}{2} (\bibinfo{year}{2023}),
  \bibinfo{pages}{283--316}.
\newblock


\bibitem[Hill et~al\mbox{.}(2016)]%
        {hill2016trials}
\bibfield{author}{\bibinfo{person}{Charles Hill}, \bibinfo{person}{Rachel
  Bellamy}, \bibinfo{person}{Thomas Erickson}, {and} \bibinfo{person}{Margaret
  Burnett}.} \bibinfo{year}{2016}\natexlab{}.
\newblock \showarticletitle{Trials and tribulations of developers of
  intelligent systems: A field study}. In \bibinfo{booktitle}{\emph{2016 IEEE
  symposium on visual languages and human-centric computing (VL/HCC)}}. IEEE,
  \bibinfo{pages}{162--170}.
\newblock


\bibitem[Hinton et~al\mbox{.}(2015)]%
        {hinton2015distilling}
\bibfield{author}{\bibinfo{person}{Geoffrey Hinton}, \bibinfo{person}{Oriol
  Vinyals}, {and} \bibinfo{person}{Jeff Dean}.}
  \bibinfo{year}{2015}\natexlab{}.
\newblock \bibinfo{title}{Distilling the Knowledge in a Neural Network}.
\newblock
\newblock
\showeprint[arxiv]{1503.02531}~[stat.ML]


\bibitem[Horkoff(2019)]%
        {8920538}
\bibfield{author}{\bibinfo{person}{Jennifer Horkoff}.}
  \bibinfo{year}{2019}\natexlab{}.
\newblock \showarticletitle{Non-Functional Requirements for Machine Learning:
  Challenges and New Directions}. In \bibinfo{booktitle}{\emph{2019 IEEE 27th
  International Requirements Engineering Conference (RE)}}.
  \bibinfo{pages}{386--391}.
\newblock
\urldef\tempurl%
\url{https://doi.org/10.1109/RE.2019.00050}
\showDOI{\tempurl}


\bibitem[Inan et~al\mbox{.}(2023)]%
        {inan2023llamaguardllmbasedinputoutput}
\bibfield{author}{\bibinfo{person}{Hakan Inan}, \bibinfo{person}{Kartikeya
  Upasani}, \bibinfo{person}{Jianfeng Chi}, \bibinfo{person}{Rashi Rungta},
  \bibinfo{person}{Krithika Iyer}, \bibinfo{person}{Yuning Mao},
  \bibinfo{person}{Michael Tontchev}, \bibinfo{person}{Qing Hu},
  \bibinfo{person}{Brian Fuller}, \bibinfo{person}{Davide Testuggine}, {and}
  \bibinfo{person}{Madian Khabsa}.} \bibinfo{year}{2023}\natexlab{}.
\newblock \bibinfo{title}{Llama Guard: LLM-based Input-Output Safeguard for
  Human-AI Conversations}.
\newblock
\newblock
\showeprint[arxiv]{2312.06674}~[cs.CL]


\bibitem[Islam et~al\mbox{.}(2024)]%
        {Islam_2024_CVPR}
\bibfield{author}{\bibinfo{person}{Khawar Islam},
  \bibinfo{person}{Muhammad~Zaigham Zaheer}, \bibinfo{person}{Arif Mahmood},
  {and} \bibinfo{person}{Karthik Nandakumar}.} \bibinfo{year}{2024}\natexlab{}.
\newblock \showarticletitle{DiffuseMix: Label-Preserving Data Augmentation with
  Diffusion Models}. In \bibinfo{booktitle}{\emph{Proceedings of the IEEE/CVF
  Conference on Computer Vision and Pattern Recognition (CVPR)}}.
  \bibinfo{pages}{27621--27630}.
\newblock


\bibitem[ISO/IEC({[n.\,d.]})]%
        {organization1991iso}
\bibfield{author}{\bibinfo{person}{ISO/IEC}.}
  \bibinfo{year}{[n.\,d.]}\natexlab{}.
\newblock \bibinfo{booktitle}{\emph{ISO/IEC 9126: Information Technology -
  Software Product Evaluation - Quality Characteristics and Guidelines for
  Their Use}}.
\newblock


\bibitem[Kochhar et~al\mbox{.}(2019)]%
        {8804445}
\bibfield{author}{\bibinfo{person}{Pavneet~Singh Kochhar}, \bibinfo{person}{Xin
  Xia}, {and} \bibinfo{person}{David Lo}.} \bibinfo{year}{2019}\natexlab{}.
\newblock \showarticletitle{Practitioners' Views on Good Software Testing
  Practices}. In \bibinfo{booktitle}{\emph{2019 IEEE/ACM 41st International
  Conference on Software Engineering: Software Engineering in Practice
  (ICSE-SEIP)}}. \bibinfo{pages}{61--70}.
\newblock
\urldef\tempurl%
\url{https://doi.org/10.1109/ICSE-SEIP.2019.00015}
\showDOI{\tempurl}


\bibitem[Kotelnikov et~al\mbox{.}(2023)]%
        {kotelnikov2023tabddpm}
\bibfield{author}{\bibinfo{person}{Akim Kotelnikov}, \bibinfo{person}{Dmitry
  Baranchuk}, \bibinfo{person}{Ivan Rubachev}, {and} \bibinfo{person}{Artem
  Babenko}.} \bibinfo{year}{2023}\natexlab{}.
\newblock \showarticletitle{TabDDPM: Modelling Tabular Data with Diffusion
  Models}. In \bibinfo{booktitle}{\emph{Proceedings of the 40th International
  Conference on Machine Learning}} \emph{(\bibinfo{series}{Proceedings of
  Machine Learning Research}, Vol.~\bibinfo{volume}{202})},
  \bibfield{editor}{\bibinfo{person}{Andreas Krause}, \bibinfo{person}{Emma
  Brunskill}, \bibinfo{person}{Kyunghyun Cho}, \bibinfo{person}{Barbara
  Engelhardt}, \bibinfo{person}{Sivan Sabato}, {and} \bibinfo{person}{Jonathan
  Scarlett}} (Eds.). \bibinfo{publisher}{PMLR}, \bibinfo{pages}{17564--17579}.
\newblock


\bibitem[Landis and Koch(1977)]%
        {landis1977measurement}
\bibfield{author}{\bibinfo{person}{J~Richard Landis} {and}
  \bibinfo{person}{Gary~G Koch}.} \bibinfo{year}{1977}\natexlab{}.
\newblock \showarticletitle{The measurement of observer agreement for
  categorical data}.
\newblock \bibinfo{journal}{\emph{biometrics}} (\bibinfo{year}{1977}),
  \bibinfo{pages}{159--174}.
\newblock


\bibitem[Lin and Gao(2022)]%
        {LIN2022118354}
\bibfield{author}{\bibinfo{person}{Kang Lin} {and} \bibinfo{person}{Yuzhuo
  Gao}.} \bibinfo{year}{2022}\natexlab{}.
\newblock \showarticletitle{Model explainability of financial fraud detection
  by group SHAP}.
\newblock \bibinfo{journal}{\emph{Expert Systems with Applications}}
  \bibinfo{volume}{210} (\bibinfo{year}{2022}), \bibinfo{pages}{118354}.
\newblock
\showISSN{0957-4174}
\urldef\tempurl%
\url{https://doi.org/10.1016/j.eswa.2022.118354}
\showDOI{\tempurl}


\bibitem[Lowell et~al\mbox{.}(2019)]%
        {lowell2019practical}
\bibfield{author}{\bibinfo{person}{David Lowell}, \bibinfo{person}{Zachary~C.
  Lipton}, {and} \bibinfo{person}{Byron~C. Wallace}.}
  \bibinfo{year}{2019}\natexlab{}.
\newblock \showarticletitle{Practical Obstacles to Deploying Active Learning}.
  In \bibinfo{booktitle}{\emph{Proceedings of the 2019 Conference on Empirical
  Methods in Natural Language Processing and the 9th International Joint
  Conference on Natural Language Processing (EMNLP-IJCNLP)}}.
  \bibinfo{publisher}{Association for Computational Linguistics},
  \bibinfo{pages}{21--30}.
\newblock
\urldef\tempurl%
\url{https://doi.org/10.18653/v1/D19-1003}
\showDOI{\tempurl}


\bibitem[Lwakatare et~al\mbox{.}(2020)]%
        {lwakatare2020large}
\bibfield{author}{\bibinfo{person}{Lucy~Ellen Lwakatare},
  \bibinfo{person}{Aiswarya Raj}, \bibinfo{person}{Ivica Crnkovic},
  \bibinfo{person}{Jan Bosch}, {and} \bibinfo{person}{Helena~Holmstr{\"o}m
  Olsson}.} \bibinfo{year}{2020}\natexlab{}.
\newblock \showarticletitle{Large-scale machine learning systems in real-world
  industrial settings: A review of challenges and solutions}.
\newblock \bibinfo{journal}{\emph{Information and software technology}}
  \bibinfo{volume}{127} (\bibinfo{year}{2020}), \bibinfo{pages}{106368}.
\newblock


\bibitem[Lyu et~al\mbox{.}(2024)]%
        {lyu2024evaluating}
\bibfield{author}{\bibinfo{person}{Yunbo Lyu}, \bibinfo{person}{Hong~Jin Kang},
  \bibinfo{person}{Ratnadira Widyasari}, \bibinfo{person}{Julia Lawall}, {and}
  \bibinfo{person}{David Lo}.} \bibinfo{year}{2024}\natexlab{}.
\newblock \showarticletitle{Evaluating szz implementations: An empirical study
  on the linux kernel}.
\newblock \bibinfo{journal}{\emph{IEEE Transactions on Software Engineering}}
  \bibinfo{volume}{50}, \bibinfo{number}{9} (\bibinfo{year}{2024}),
  \bibinfo{pages}{2219--2239}.
\newblock


\bibitem[Lyu et~al\mbox{.}(2023)]%
        {lyu2023chronos}
\bibfield{author}{\bibinfo{person}{Yunbo Lyu}, \bibinfo{person}{Thanh Le-Cong},
  \bibinfo{person}{Hong~Jin Kang}, \bibinfo{person}{Ratnadira Widyasari},
  \bibinfo{person}{Zhipeng Zhao}, \bibinfo{person}{Xuan-Bach~D Le},
  \bibinfo{person}{Ming Li}, {and} \bibinfo{person}{David Lo}.}
  \bibinfo{year}{2023}\natexlab{}.
\newblock \showarticletitle{Chronos: Time-aware zero-shot identification of
  libraries from vulnerability reports}. In \bibinfo{booktitle}{\emph{2023
  IEEE/ACM 45th International Conference on Software Engineering (ICSE)}}.
  IEEE, \bibinfo{pages}{1033--1045}.
\newblock


\bibitem[Lyu et~al\mbox{.}(2025a)]%
        {lyu2025existing}
\bibfield{author}{\bibinfo{person}{Yunbo Lyu}, \bibinfo{person}{Zhou Yang},
  \bibinfo{person}{Yuqing Niu}, \bibinfo{person}{Jing Jiang}, {and}
  \bibinfo{person}{David Lo}.} \bibinfo{year}{2025}\natexlab{a}.
\newblock \showarticletitle{Do existing testing tools really uncover gender
  bias in text-to-image models?}. In \bibinfo{booktitle}{\emph{Proceedings of
  the 33rd ACM International Conference on Multimedia}}.
  \bibinfo{pages}{11687--11696}.
\newblock


\bibitem[Lyu et~al\mbox{.}(2025b)]%
        {lyu2025my}
\bibfield{author}{\bibinfo{person}{Yunbo Lyu}, \bibinfo{person}{Zhou Yang},
  \bibinfo{person}{Jieke Shi}, \bibinfo{person}{Jianming Chang},
  \bibinfo{person}{Yue Liu}, {and} \bibinfo{person}{David Lo}.}
  \bibinfo{year}{2025}\natexlab{b}.
\newblock \showarticletitle{" My productivity is boosted, but…" Demystifying
  Users’ Perception on AI Coding Assistants}. In
  \bibinfo{booktitle}{\emph{2025 40th IEEE/ACM International Conference on
  Automated Software Engineering (ASE)}}. IEEE, \bibinfo{pages}{191--203}.
\newblock


\bibitem[Nahar et~al\mbox{.}(2025)]%
        {nahar2025beyond}
\bibfield{author}{\bibinfo{person}{Nadia Nahar}, \bibinfo{person}{Christian
  Kastner}, \bibinfo{person}{Jenna Butler}, \bibinfo{person}{Chris Parnin},
  \bibinfo{person}{Thomas Zimmermann.}, {and} \bibinfo{person}{Christian
  Bird}.} \bibinfo{year}{2025}\natexlab{}.
\newblock \showarticletitle{{ Beyond the Comfort Zone: Emerging Solutions to
  Overcome Challenges in Integrating LLMs into Software Products }}. In
  \bibinfo{booktitle}{\emph{2025 IEEE/ACM 47th International Conference on
  Software Engineering: Software Engineering in Practice (ICSE-SEIP)}}.
  \bibinfo{publisher}{IEEE Computer Society}, \bibinfo{address}{Los Alamitos,
  CA, USA}, \bibinfo{pages}{516--527}.
\newblock
\urldef\tempurl%
\url{https://doi.org/10.1109/ICSE-SEIP66354.2025.00051}
\showDOI{\tempurl}


\bibitem[Nahar et~al\mbox{.}(2023)]%
        {nahar2023meta}
\bibfield{author}{\bibinfo{person}{Nadia Nahar}, \bibinfo{person}{Haoran
  Zhang}, \bibinfo{person}{Grace Lewis}, \bibinfo{person}{Shurui Zhou}, {and}
  \bibinfo{person}{Christian K{\"a}stner}.} \bibinfo{year}{2023}\natexlab{}.
\newblock \showarticletitle{A meta-summary of challenges in building products
  with ml components--collecting experiences from 4758+ practitioners}. In
  \bibinfo{booktitle}{\emph{2023 IEEE/ACM 2nd International Conference on AI
  Engineering--Software Engineering for AI (CAIN)}}. IEEE,
  \bibinfo{pages}{171--183}.
\newblock


\bibitem[Newman(2021)]%
        {newman2021building}
\bibfield{author}{\bibinfo{person}{Sam Newman}.}
  \bibinfo{year}{2021}\natexlab{}.
\newblock \bibinfo{booktitle}{\emph{Building microservices: designing
  fine-grained systems}}.
\newblock \bibinfo{publisher}{" O'Reilly Media, Inc."}.
\newblock


\bibitem[of~the Comptroller of~the Currency(2025)]%
        {occ_sar}
\bibfield{author}{\bibinfo{person}{Office of~the Comptroller of~the Currency}.}
  \bibinfo{year}{2025}\natexlab{}.
\newblock \bibinfo{title}{Suspicious Activity Reports (SAR)}.
\newblock
\newblock
\urldef\tempurl%
\url{https://www.occ.treas.gov/topics/supervision-and-examination/bank-operations/financial-crime/suspicious-activity-reports/index-suspicious-activity-reports.html}
\showURL{%
\tempurl}


\bibitem[Paleyes et~al\mbox{.}(2022)]%
        {10.1145/3533378}
\bibfield{author}{\bibinfo{person}{Andrei Paleyes},
  \bibinfo{person}{Raoul-Gabriel Urma}, {and} \bibinfo{person}{Neil~D.
  Lawrence}.} \bibinfo{year}{2022}\natexlab{}.
\newblock \showarticletitle{Challenges in Deploying Machine Learning: A Survey
  of Case Studies}.
\newblock \bibinfo{journal}{\emph{ACM Comput. Surv.}} \bibinfo{volume}{55},
  \bibinfo{number}{6}, Article \bibinfo{articleno}{114} (\bibinfo{date}{dec}
  \bibinfo{year}{2022}), \bibinfo{numpages}{29}~pages.
\newblock
\showISSN{0360-0300}
\urldef\tempurl%
\url{https://doi.org/10.1145/3533378}
\showDOI{\tempurl}


\bibitem[Salama et~al\mbox{.}(2019)]%
        {salama2019pruning}
\bibfield{author}{\bibinfo{person}{Abdullah Salama}, \bibinfo{person}{Oleksiy
  Ostapenko}, \bibinfo{person}{Tassilo Klein}, {and} \bibinfo{person}{Moin
  Nabi}.} \bibinfo{year}{2019}\natexlab{}.
\newblock \bibinfo{title}{Pruning at a Glance: Global Neural Pruning for Model
  Compression}.
\newblock
\newblock
\showeprint[arxiv]{1912.00200}~[cs.CV]


\bibitem[Serban et~al\mbox{.}(2020)]%
        {10.1145/3382494.3410681}
\bibfield{author}{\bibinfo{person}{Alex Serban}, \bibinfo{person}{Koen van~der
  Blom}, \bibinfo{person}{Holger Hoos}, {and} \bibinfo{person}{Joost Visser}.}
  \bibinfo{year}{2020}\natexlab{}.
\newblock \showarticletitle{Adoption and Effects of Software Engineering Best
  Practices in Machine Learning}. In \bibinfo{booktitle}{\emph{Proceedings of
  the 14th ACM / IEEE International Symposium on Empirical Software Engineering
  and Measurement (ESEM)}} (Bari, Italy) \emph{(\bibinfo{series}{ESEM '20})}.
  \bibinfo{publisher}{Association for Computing Machinery},
  \bibinfo{address}{New York, NY, USA}, Article \bibinfo{articleno}{3},
  \bibinfo{numpages}{12}~pages.
\newblock
\showISBNx{9781450375801}
\urldef\tempurl%
\url{https://doi.org/10.1145/3382494.3410681}
\showDOI{\tempurl}


\bibitem[Serban et~al\mbox{.}(2024)]%
        {serban2024software}
\bibfield{author}{\bibinfo{person}{Alex Serban}, \bibinfo{person}{Koen van~der
  Blom}, \bibinfo{person}{Holger Hoos}, {and} \bibinfo{person}{Joost Visser}.}
  \bibinfo{year}{2024}\natexlab{}.
\newblock \showarticletitle{Software engineering practices for machine
  learning—Adoption, effects, and team assessment}.
\newblock \bibinfo{journal}{\emph{Journal of Systems and Software}}
  \bibinfo{volume}{209} (\bibinfo{year}{2024}), \bibinfo{pages}{111907}.
\newblock


\bibitem[Siebert et~al\mbox{.}(2022)]%
        {siebert2022construction}
\bibfield{author}{\bibinfo{person}{Julien Siebert}, \bibinfo{person}{Lisa
  Joeckel}, \bibinfo{person}{Jens Heidrich}, \bibinfo{person}{Adam Trendowicz},
  \bibinfo{person}{Koji Nakamichi}, \bibinfo{person}{Kyoko Ohashi},
  \bibinfo{person}{Isao Namba}, \bibinfo{person}{Rieko Yamamoto}, {and}
  \bibinfo{person}{Mikio Aoyama}.} \bibinfo{year}{2022}\natexlab{}.
\newblock \showarticletitle{Construction of a quality model for machine
  learning systems}.
\newblock \bibinfo{journal}{\emph{Software Quality Journal}}
  \bibinfo{volume}{30}, \bibinfo{number}{2} (\bibinfo{year}{2022}),
  \bibinfo{pages}{307--335}.
\newblock


\bibitem[Sinha and Lee(2024)]%
        {sinha2024challenges}
\bibfield{author}{\bibinfo{person}{Sudhi Sinha} {and} \bibinfo{person}{Young~M
  Lee}.} \bibinfo{year}{2024}\natexlab{}.
\newblock \showarticletitle{Challenges with developing and deploying AI models
  and applications in industrial systems}.
\newblock \bibinfo{journal}{\emph{Discover Artificial Intelligence}}
  \bibinfo{volume}{4}, \bibinfo{number}{1} (\bibinfo{year}{2024}),
  \bibinfo{pages}{55}.
\newblock


\bibitem[Song et~al\mbox{.}(2022)]%
        {song2022exploring}
\bibfield{author}{\bibinfo{person}{Qunying Song}, \bibinfo{person}{Markus
  Borg}, \bibinfo{person}{Emelie Engström}, \bibinfo{person}{Håkan Ardö},
  {and} \bibinfo{person}{Sergio Rico}.} \bibinfo{year}{2022}\natexlab{}.
\newblock \bibinfo{title}{Exploring ML testing in practice -- Lessons learned
  from an interactive rapid review with Axis Communications}.
\newblock
\newblock
\showeprint[arxiv]{2203.16225}~[cs.SE]


\bibitem[Villamizar and Kalinowski(2024)]%
        {villamizar2024identifying}
\bibfield{author}{\bibinfo{person}{Hugo Villamizar} {and}
  \bibinfo{person}{Marcos Kalinowski}.} \bibinfo{year}{2024}\natexlab{}.
\newblock \showarticletitle{Identifying concerns when specifying machine
  learning-enabled systems: A perspective-based approach}. In
  \bibinfo{booktitle}{\emph{Proceedings of the XXIII Brazilian Symposium on
  Software Quality}}. \bibinfo{pages}{673--675}.
\newblock


\bibitem[Widyassari et~al\mbox{.}(2022)]%
        {widyassari2022review}
\bibfield{author}{\bibinfo{person}{Adhika~Pramita Widyassari},
  \bibinfo{person}{Supriadi Rustad}, \bibinfo{person}{Guruh~Fajar Shidik},
  \bibinfo{person}{Edi Noersasongko}, \bibinfo{person}{Abdul Syukur},
  \bibinfo{person}{Affandy Affandy}, {and} \bibinfo{person}{De~Rosal
  Ignatius~Moses Setiadi}.} \bibinfo{year}{2022}\natexlab{}.
\newblock \showarticletitle{Review of automatic text summarization techniques
  \& methods}.
\newblock \bibinfo{journal}{\emph{Journal of King Saud University-Computer and
  Information Sciences}} \bibinfo{volume}{34}, \bibinfo{number}{4}
  (\bibinfo{year}{2022}), \bibinfo{pages}{1029--1046}.
\newblock


\bibitem[Wong et~al\mbox{.}(2021)]%
        {wong2021external}
\bibfield{author}{\bibinfo{person}{Andrew Wong}, \bibinfo{person}{Erkin Otles},
  \bibinfo{person}{John~P Donnelly}, \bibinfo{person}{Andrew Krumm},
  \bibinfo{person}{Jeffrey McCullough}, \bibinfo{person}{Olivia
  DeTroyer-Cooley}, \bibinfo{person}{Justin Pestrue}, \bibinfo{person}{Marie
  Phillips}, \bibinfo{person}{Judy Konye}, \bibinfo{person}{Carleen Penoza},
  {et~al\mbox{.}}} \bibinfo{year}{2021}\natexlab{}.
\newblock \showarticletitle{External validation of a widely implemented
  proprietary sepsis prediction model in hospitalized patients}.
\newblock \bibinfo{journal}{\emph{JAMA internal medicine}}
  \bibinfo{volume}{181}, \bibinfo{number}{8} (\bibinfo{year}{2021}),
  \bibinfo{pages}{1065--1070}.
\newblock


\bibitem[Xiangyu et~al\mbox{.}(2017)]%
        {xiangyu2017deep}
\bibfield{author}{\bibinfo{person}{Zhao Xiangyu}, \bibinfo{person}{Zhang
  Liang}, \bibinfo{person}{Xia Long}, \bibinfo{person}{Ding Zhuoye},
  \bibinfo{person}{Yin Dawei}, {and} \bibinfo{person}{Tang Jiliang}.}
  \bibinfo{year}{2017}\natexlab{}.
\newblock \showarticletitle{Deep Reinforcement Learning for List-wise
  Recommendations}.
\newblock \bibinfo{journal}{\emph{arXiv preprint arXiv:1801.00209}}
  (\bibinfo{year}{2017}).
\newblock
\urldef\tempurl%
\url{https://www.arxiv.org/abs/1801.00209}
\showURL{%
\tempurl}


\bibitem[Yang et~al\mbox{.}(2022)]%
        {yang2022revisiting}
\bibfield{author}{\bibinfo{person}{Zhou Yang}, \bibinfo{person}{Jieke Shi},
  \bibinfo{person}{Muhammad~Hilmi Asyrofi}, {and} \bibinfo{person}{David Lo}.}
  \bibinfo{year}{2022}\natexlab{}.
\newblock \showarticletitle{Revisiting Neuron Coverage Metrics and Quality of
  Deep Neural Networks}. In \bibinfo{booktitle}{\emph{2022 IEEE International
  Conference on Software Analysis, Evolution and Reengineering (SANER)}}.
  \bibinfo{publisher}{IEEE Computer Society}, \bibinfo{address}{Los Alamitos,
  CA, USA}, \bibinfo{pages}{408--419}.
\newblock
\showISSN{1534-5351}


\bibitem[You et~al\mbox{.}(2025)]%
        {you2025navigating}
\bibfield{author}{\bibinfo{person}{Hanmo You}, \bibinfo{person}{Zan Wang},
  \bibinfo{person}{Bin Lin}, {and} \bibinfo{person}{Junjie Chen}.}
  \bibinfo{year}{2025}\natexlab{}.
\newblock \showarticletitle{{ Navigating the Testing of Evolving Deep Learning
  Systems: An Exploratory Interview Study }}. In \bibinfo{booktitle}{\emph{2025
  IEEE/ACM 47th International Conference on Software Engineering (ICSE)}}.
  \bibinfo{publisher}{IEEE Computer Society}, \bibinfo{address}{Los Alamitos,
  CA, USA}, \bibinfo{pages}{643--643}.
\newblock
\showISSN{1558-1225}
\urldef\tempurl%
\url{https://doi.org/10.1109/ICSE55347.2025.00106}
\showDOI{\tempurl}


\bibitem[Zhang et~al\mbox{.}(2022)]%
        {zhang2019machine}
\bibfield{author}{\bibinfo{person}{Jie~M. Zhang}, \bibinfo{person}{Mark
  Harman}, \bibinfo{person}{Lei Ma}, {and} \bibinfo{person}{Yang Liu}.}
  \bibinfo{year}{2022}\natexlab{}.
\newblock \showarticletitle{Machine Learning Testing: Survey, Landscapes and
  Horizons}.
\newblock \bibinfo{journal}{\emph{IEEE Transactions on Software Engineering}}
  \bibinfo{volume}{48}, \bibinfo{number}{1} (\bibinfo{year}{2022}),
  \bibinfo{pages}{1--36}.
\newblock
\urldef\tempurl%
\url{https://doi.org/10.1109/TSE.2019.2962027}
\showDOI{\tempurl}


\bibitem[Zhang et~al\mbox{.}(2019)]%
        {zhang2019software}
\bibfield{author}{\bibinfo{person}{Xufan Zhang}, \bibinfo{person}{Yilin Yang},
  \bibinfo{person}{Yang Feng}, {and} \bibinfo{person}{Zhenyu Chen}.}
  \bibinfo{year}{2019}\natexlab{}.
\newblock \showarticletitle{Software engineering practice in the development of
  deep learning applications}.
\newblock \bibinfo{journal}{\emph{arXiv preprint arXiv:1910.03156}}
  (\bibinfo{year}{2019}).
\newblock


\end{thebibliography}

\end{document}